\documentclass[11pt, a4paper]{article}
\pdfoutput=1

\usepackage{jheppub}
\usepackage{lipsum}
\usepackage[all]{hypcap}
\usepackage{multirow}
\usepackage{titlesec}
\titleformat*{\section}{\bfseries\boldmath}
\titleformat*{\subsection}{\bfseries\boldmath}
\titleformat*{\subsubsection}{\bfseries\boldmath}
\usepackage[detect-all]{siunitx}
\usepackage{hyperref} 

\newcommand{\br}{\textrm{BR}}
\newcommand{\sm}{\textrm{SM}}
\newcommand{\np}{\textrm{NP}}
\newcommand{\unt}{\textrm{unt}}
\newcommand{\inv}{\textrm{inv}}
\newcommand{\tev}{\,\textrm{TeV}}
\newcommand{\gev}{\,\textrm{GeV}}
\newcommand{\mev}{\,\textrm{MeV}}

\newcommand{\st}{\sin^2\theta}
\newcommand{\sint}{\textrm{s}_{\theta}}
\newcommand{\sw}{\ensuremath{\sin\theta_\textnormal{w}}}
\newcommand{\cw}{\ensuremath{\cos\theta_\textnormal{w}}}
\newcommand{\swsq}{\ensuremath{\sin^2\theta_\textnormal{w}}}
\newcommand{\cwsq}{\ensuremath{\cos^2\theta_\textnormal{w}}}
\newcommand{\swquart}{\ensuremath{\sin^4\theta_\textnormal{w}}}
\newcommand{\cwquart}{\ensuremath{\cos^4\theta_\textnormal{w}}}

\newcommand{\mphi}{\ensuremath{m_{\phi}}}
\newcommand{\mphisq}{\ensuremath{\mphi^2}}
\newcommand{\mZ}{\ensuremath{m_Z}}
\newcommand{\mZsq}{\ensuremath{m_Z^2}}

\newcommand{\lint}{\mathcal{L}_{\textrm{int}} }

\newcommand{\fb}{\,\textrm{fb} }
\newcommand{\ab}{\,\textrm{ab} }
\newcommand{\iab}{\,\ab^{-1}}
\newcommand{\ifb}{\,\fb^{-1}}
\newcommand{\pb}{\,\textrm{pb} }

\newcommand{\cp}{\ensuremath{\mathcal{CP}}}
\newcommand{\rbr}{r_{\rm br} }
\newcommand{\sth}{s_{\theta}}
\newcommand{\cth}{c_{\theta}}
\newcommand{\dkaZ}{\delta_{\kappa_Z}}

\title{\large Relaxion and light (pseudo)scalars at the HL-LHC and lepton colliders}
\author{Claudia Frugiuele,}
\author{Elina Fuchs,}
\author{Gilad Perez,}
\author{Matthias Schlaffer}

\affiliation{Weizmann Institute of Science\\234 Herzl Street\\Rehovot 7610001\\Israel}

\emailAdd{Claudia.Frugiuele@weizmann.ac.il}
\emailAdd{Elina.Fuchs@weizmann.ac.il}
\emailAdd{Gilad.Perez@weizmann.ac.il}
\emailAdd{Matthias.Schlaffer@weizmann.ac.il}

\abstract{ We study the potential of future lepton colliders, running at the $Z$-pole and above, and
  the High-Luminosity LHC to search for the relaxion and other light scalars $\phi$.  We investigate
  the interplay of direct searches and precision observables for both \cp{}-even and -odd couplings.
  In particular, precision measurements of exotic $Z$-decays, Higgs couplings, the exotic Higgs
  decay into a relaxion pair and associated $Z\phi$ and $\gamma \phi$ production are promising
  channels to yield strong bounds.}

\begin{document}

\maketitle

\section{Introduction}
\label{sec:introduction}
The discovery of the Higgs boson in 2012 was an extraordinary success for both the Standard Model
(SM) and the Large Hadron Collider (LHC).  Six years later, the LHC is forcing us to reconsider our
views regarding the existence of New Physics (NP) at or near the electroweak scale.  Despite a rich
program of dedicated searches in many channels, no conclusive signs of NP between the electroweak
and the TeV scale have been found.  Consequently, the idea is questioned that NP solving the
hierarchy problem is at the few-TeV scale and traditional symmetry-based approaches are being
challenged.

Generally speaking, constraints on new particles coupled to the SM via either the gauge interactions
or other couplings of similar strength are getting increasingly strong, and push models with such
couplings to non-generic corners of parameter space.  As a consequence, a growing attention is now
given to an alternative point of view where the additional degrees of freedom are light and have
ultra-weak couplings to the SM, and hence are hard to find.  Searching for NP of this kind can be
motivated based on purely phenomenological reasonings, but it is also motivated by generic
theoretical concepts that support the existence of new weakly coupled light fields.  Such moduli or
pseudo-Nambu-Goldstone-Bosons (pNGBs) appear in many extensions of the SM, see
\textit{e.g.}~Ref.~\cite{mina} for a recent review and references therein.  The possibility for the
presence of these new light fields is further supported by theoretical frameworks that address other
conceptual and observational open questions such as the hierarchy and the strong \cp{} problems or
the origin of dark matter (see \textit{e.g.}~Refs.~\cite{Peccei:1977hh, Davidi:2017gir,
  Davidi:2018sii, Hu:2000ke, Wilczek:1982rv}).

Recently, a new mechanism has been proposed that addresses the hierarchy problem in a way that goes
beyond the conventional paradigm of a symmetry-based solution to fine-tuning.  It belongs to the
abovementioned class of models where the solution is associated with the existence of a new and
special kind of pNGB, and it is worth to be examined in detail.  It is denoted as the relaxion
mechanism~\cite{kaplan}, where the pNGB---the relaxion---stabilizes the Higgs mass dynamically. The
Higgs mass depends on the classical relaxion field value which evolves in time. The relaxion rolls
down a potential, eventually stopping its rolling at a special field value where the Higgs mass is
much smaller than the theory's cutoff, hence addressing the fine-tuning problem. Since the rolling
needs to be very slow, a frictional force is necessary.  The main sources of friction that were
proposed thus far are either the Hubble friction during inflation, thereby linking the solution of
the hierarchy problem to cosmology~\cite{kaplan}, or alternatively through particle production, see
Ref.~\cite{Hook:2016mqo}. Relaxion models do not require TeV-scale top, gauge or Higgs
partners. However, they do generically lead to an interesting phenomenology. In addition to
cosmological signatures~\cite{Fonseca:2018xzp}, relaxion models can also leave fingerprints at the
low energy precision frontier, the intensity frontier and colliders~\cite{Flacke:2016szy,choiim}.

As already discussed in Refs.~\cite{Flacke:2016szy,choiim}, the relaxion parameter space spans many
orders of magnitude in mass and coupling, see Fig.~\ref{fig:summary}.  This is somewhat similar to
the case of the axion, but also different in ways discussed below.  The relaxion can be as light as
$10^{-20}$ eV (\textit{e.g.}~for Hubble-friction based models~\cite{kaplan}, and ignoring the
relaxion's quality problem~\cite{Davidi:2017gir}) up to tens of
GeV~\cite{Flacke:2016szy,Davidi:2017gir}. This is a huge parameter space to probe and it cannot be
scanned by a single experiment, or a single frontier of new physics searches. Even more so, most
experiments do not even touch the physical parameter space of relaxion models. It is left to
consider which of the High-Luminosity (HL)-LHC and future lepton colliders have the power to probe
relaxion dynamics.

In this paper we mostly focus on searching for GeV-scale relaxions, which are ultra-heavy with
respect to the relaxion framework, but at the same time at the low-mass end of the collider reach
and therefore pose an experimental challenge.  Admittedly, this region covers only a small part of
the parameter space of the framework, however, as we shall see, future experiments will have the
power to significantly probe physical regions of it. Furthermore, this region, which addresses the
little hierarchy problem~\cite{familon}, is typically free of some of the model's theoretical flaws
such as an exponential number of e-folds~\cite{kaplan,Patil:2015oxa}, the need for significant
clockworking~\cite{cky,kaplan2,Giudice:2016yja}, and finally a quality
problem~\cite{Davidi:2017gir,Gupta:2018wif}.

As for the present, low-mass resonances are already being searched for at the LHC, but the further
development of reconstruction and analysis techniques could boost its power in this regime.  Indeed,
an increased effort has been invested to show that the LHC is powerful in probing several
interesting BSM scenarios featuring light particles~\cite{Sirunyan:2017nvi, Cacciapaglia:2017iws,
  Jaeckel:2015jla, Curtin:2014cca, Haisch:2018kqx, Mariotti:2017vtv, Bellazzini:2017neg}.

We focus in particular on an interesting property of the relaxion, namely that it mixes with the
Higgs due to the fact that \cp{} is spontaneously broken by its vacuum expectation value (vev), see
\textit{e.g.} Refs.~\cite{Flacke:2016szy, Davidi:2017gir}.  The presence of a light new scalar mixed
with the SM-like Higgs is not only a feature common to several new approaches to solve the hierarchy
problem~\cite{kaplan,Arkani-Hamed:2016rle,Hook:2018jle}, but it is ubiquitous in many other more
traditional scenarios for physics beyond the SM (BSM) such as the singlet
extension~\cite{Robens:2016xkb}, the Higgs portal~\cite{Falkowski:2015iwa,Batell:2016ove}, 2-Higgs
doublet models (2HDM)~\cite{Cacciapaglia:2016tlr,Chun:2017yob,Ghosh:2015gpa}, supersymmetry
(SUSY)~\cite{Ellwanger:2015uaz,Bertuzzo:2014bwa,Bellazzini:2017neg} and several others.

As a theoretically motivated framework with a light pNGB with both \cp{}-even and -odd couplings,
the relaxion provides us with a useful benchmark model to examine the capabilities of the HL-LHC and
future lepton colliders to probe the corresponding parameter space in direct and indirect channels.
While our discussion mainly focuses on the relaxion framework, most of the resulting bounds also
apply to the singlet extension, Higgs portal and other models.

Our paper is structured as follows. In Sect.~\ref{sec:relaxion-revisited} we summarize the relaxion
mechanism and highlight its couplings that are relevant for collider phenomenology. In addition, we
collect in Sect.~\ref{sec:stat-exper-prob} the existing bounds on the model's rather wide parameter
space.  In Sect.~\ref{sec:prospecs-at-high} we discuss in detail direct and indirect probes of
\cp{}-even and -odd couplings, both at hadron and lepton colliders.  Finally we conclude in
Sect.~\ref{sec:conclusion}.

\section{Relaxion phenomenology}
\label{sec:relaxion-revisited}

In the following we will briefly review the relaxion mechanism. The effective scalar potential of
the theory depends both on the Higgs doublet $H$ and the relaxion $\phi$,
\begin{align}
  V(H,\phi)&=\mu^2(\phi) H^\dagger H+\lambda (H^ \dagger H)^2 + V_{\rm sr}(\phi)+ V_\text{br}(h,\phi)\, ,\label{pot}\\
  \mu^2(\phi)&=-\Lambda^2+g \Lambda \phi + \dots   \, ,
\label{mu2}
\end{align}
where $h$ is the physical component of the Higgs doublet (before mixing with $\phi$, see below), and
$\Lambda$ is the cutoff scale of a Higgs loop and. As discussed in Ref.~\cite{kaplan}, $\Lambda $
needs to be significantly smaller than the Planck mass. Hence at energies above $ \Lambda$,
additional structure is required to fully stabilize the Higgs mass. See Refs.~\cite{Batell:2015fma,
  Evans:2016htp, Batell:2017kho} for attempts in UV completing relaxion models by supersymmetry and
composite Higgs.  During its evolution, the relaxion scans the Higgs mass parameter $\mu^2(\phi)$
from a large and positive cutoff energy $\Lambda^2$ down to negative values because of the slow-roll
potential
\begin{equation}
  V_{\rm sr}(\phi)=r g\Lambda^3 \phi\,,
\label{roll}
\end{equation}
where $g$ is a (small) dimensionless coupling which breaks the shift symmetry of the backreaction
potential $V_\text{br}$ and $r > \frac{1} {16 \pi^2}$ due to naturalness requirements. Once
$\mu^2(\phi)$ becomes negative, the Higgs gets a vev $v^2(\phi)=-\frac{\mu^2(\phi)}{\lambda}$. This
non-zero vev activates a backreaction potential $V_\text{br}$ which eventually stops the rolling of
the relaxion at a value $\phi_0$, where $v(\phi_0)=246\gev$.

The properties of the backreaction mechanism are model-dependent. In the minimal relaxion model
discussed in Ref.~\cite{kaplan}, $V_{\rm br}$ is generated by low-energy QCD and thus the relaxion
is identified with the QCD axion. Yet, this setup typically predicts a too large phase
$\theta_\text{QCD}$ and is therefore ruled out by the upper bound on the neutron electric dipole
moment. Possible ways to suppress the $\cp$ violation associated with the relaxion mechanism are
discussed in Refs.~\cite{kaplan,Nelson:2017cfv}.  Alternatively one can introduce a new sector with
strongly~\cite{kaplan} or weakly~\cite{familon} coupled vector-like fermions, which generates a
Higgs-dependent backreaction potential of the form
\begin{eqnarray}
  V_\text{br}( h,\phi)= -\tilde M^{4-j} \left(\frac{v(\phi)+h}{\sqrt{2}}\right)^j \cos \left( \frac{\phi}{f} \right),
   \label{eq:br}
  \end{eqnarray} 
  with $1\leq j\leq 4$ and $\tilde M$ being a mass parameter~\cite{Flacke:2016szy}. For later
  convenience we define the backreaction scale
  $\Lambda_\text{br}(v(\phi))^4 \equiv \tilde M^{4-j} v(\phi)^j/\sqrt{2}^j$.  This scale is not
  predicted. It can be close to (or larger than) than the electroweak scale or as small as sub-GeV
  as in the QCD axion case~\cite{kaplan}, which corresponds to $j=1$.  In this paper we consider
  only the case of $j=2$.  In order to avoid fine-tuning, the backreaction scale is limited
  by~\cite{Espinosa:2015eda}
 \begin{equation}
 \Lambda_{\rm br}(v(\phi_0))^2 \lesssim  \mathcal{O}(2 \pi v^2)
 \label{lambdabrMAX}\,.
\end{equation}

In the following we will focus on the class of relaxion models where the backreaction has the form
of Eq.~\eqref{eq:br} and investigate its phenomenological consequences. Other realizations of the
relaxion mechanism have been introduced for instance in
Refs.~\cite{Hook:2016mqo,Espinosa:2015eda,Nelson:2017cfv}, where our analysis does not apply.

\subsection{Relevant parameters}
\label{sec:RHmix}
In Refs.~\cite{Flacke:2016szy,Choi:2016luu} the couplings of the relaxion to the SM particles and
their phenomenological consequences were studied.  It was pointed out that generically the
backreaction and/or the relaxion vev~\cite{Davidi:2017gir} lead to \cp{} violation. As a result the
relaxion mixes with the Higgs and inherits its couplings to SM fields.

It is convenient to write
\begin{equation}
\Lambda_{\rm br} = r_{\rm br} v\,,
\end{equation}
where $r_{\rm br} \lesssim \sqrt{2\pi}$ from Eq.~\eqref{lambdabrMAX}.

From the diagonalization of the $2\times 2$ mass matrix (see Ref.~\cite{Flacke:2016szy} for the full
expressions) and imposing the largest eigenvalue to be the Higgs mass of $m_h=125\gev$ we obtain
\begin{align}
  m^2_{\phi}
  &= \frac{ \rbr^4 v^4\,\left[f^2 \left(c_0 m_h^2 - 4\rbr^4 v^2 s_0^2\right) - c_0^2 \rbr^4 v^4 \right]}{f^2\,\left(m_h^2 f^2 -c_0 \rbr^4 v^4\right)} \label{eq:exactMh}\\
  &\simeq\frac{ r_{\rm br}^4 v^4 \left[f^2 (c_0 - 16  r_{\rm br}^4 s_0^2 ) - 
    4 c_0^2  r_{\rm br}^4 v^2\right] }{f^2\,\left(f^2 - 4  c_0  r_{\rm br}^4 v^2\right)}\,,
 \label{eq:exact}
\end{align}
where the approximation holds for $m_h\approx v/2$. Here and in the following we denote
$s_0\equiv\sin(\phi_0/f)$, $c_0\equiv\cos(\phi_0/f)$, where $\phi_0$ is the endpoint of the rolling
of the relaxion. Moreover, to keep the notation simple, we also denote the mixed mass eigenstates by
$h$ and $\phi$. In the limit of $ f \gg r^2_{\rm br} v$, the expression for the relaxion mass is
simplified to
\begin{equation}
  \label{eq:relaxionSimplifiedMass}
  m_{\phi} \simeq \frac{ r_{\rm br}^2 v^2 }{f} \sqrt{c_0 -16 r_{\rm br}^4 s_0^2 }\,.  
\end{equation}
This limit requires $16\rbr^4 s_0^2 < c_0$ which demands either a small backreaction scale or a
suppressed $\cp$-violating angle $s_0$.  This condition is easily met since for
$\Lambda_{\rm br} > v$ the end-point of the relaxion rolling has a suppressed $s_0$ as clarified in
Ref.~\cite{Choi:2016luu}.  The mixing with the Higgs is given by
\begin{equation}
\sin{\theta}= \left|\frac{ 8 f r_{\rm br}^4  s_0 v}{ \sqrt{
64 f^2 r_{\rm br}^8 s_0^2 v^2 + (f^2 - 4  c_0 r_{\rm br}^4 v^2)^2}}\right|
 \simeq \left| 8   r_{\rm br}^4 s_0 \frac{v}{ f  }\right|\,,
\label{eq:sinth}
\end{equation}
where the approximation holds in the limit of $ f \gg r^2_{\rm br} v$. For later use we define
$s_\theta\equiv \sin\theta$ and $c_\theta\equiv \cos\theta$. When $16\,\rbr^4 s_0^2 \ll c_0$,
Eq.~(\ref{eq:relaxionSimplifiedMass}) further simplifies to
\begin{equation}
m_{\phi} \simeq \frac{ r_{\rm br}^2 v^2 }{f} \sqrt{c_0 }\,.
\label{eq:mphiApprox}
\end{equation}
Moreover, using Eqs.~\eqref{eq:sinth} and \eqref{eq:mphiApprox} we find that the mixing is bounded
by
\begin{equation}
 \sin\theta \leq 2\frac{\mphi}{v}.
 \label{eq:maxmixNatural}
\end{equation}
This bound holds within the above approximations, which are fulfilled in most of the parameter
space.

\subsection{The relaxion-Higgs sector}
\label{sec:RHiggs}

As discussed in Ref.~\cite{Flacke:2016szy}, via its mixing with the Higgs, the relaxion inherits the
Higgs couplings to SM particles $g_{hX}$, where $X=f,V$ suppressed by a universal factor
\textit{i.e.}~the mixing angle $\sin \theta$. This is precisely the case as for Higgs-portal models.
The couplings of the relaxion $\phi$ to fermions $f$ and vector bosons $V$ are thus given by
\begin{eqnarray}
g_{\phi X} = \sin \theta \,g_{h X} \,. \label{eq:gphipsi}
\end{eqnarray}
The upper bound on the Higgs-relaxion mixing is of the same strength as in general Higgs portal
models (see \textit{e.g.}~Refs.~\cite{pospelov,Arvanitaki:2014faa}), where the mixing of $\phi$ with
the Higgs as a function of the mass $\mphi$ is bounded by naturalness as
$ \sin\theta \leq 2\frac{\mphi}{v}$, \textit{cf.}~Eq.~(\ref{eq:maxmixNatural}).

Besides the couplings to SM fermions and gauge bosons, the relaxion also couples to the Higgs boson
via $c_{h\phi\phi}\, h\phi\phi$ which is given by (see Ref.~\cite{Flacke:2016szy}, but here with a
general $s_0, c_0$)
\begin{equation}
c_{\phi\phi h} = \frac{\rbr^4 v^3}{f^2} c_0 \cth^3
- \frac{2 \rbr^4  v^2}{f} s_0 \cth^2 \sth
- \frac{\rbr^4 v^4}{2f^3} s_0 \cth^2 \sth
- \frac{2\rbr^4 v^3}{f^2} c_0 \cth \sth^2
+ 3v \lambda \cth \sth^2
+ \frac{\rbr^4 v^2}{f} s_0 \sth^3\,.
 \label{eq:phiphih}
\end{equation}
Note that only two of the parameters $f$, $r_\textrm{br}$ and $\theta$ are independent. In the limit
of small relaxion-Higgs mixing, \textit{i.e.}~$\sin\theta \to0$, the coupling $c_{\phi\phi h}$ does
not vanish, but is reduced to the first term with $\cth\to1$, originating from the backreaction
potential.  Furthermore, the comparison with the expression for $\mphi$ in Eq.~\eqref{eq:mphiApprox}
allows for $c_0\gg 16\rbr^4 s_0^2$ to express the coupling in terms of the mass as
\begin{equation}
  c_{\phi\phi h}|_{\theta\to0} \simeq \frac{\rbr^4 v^3}{f^2} c_0 \cth^3
  \simeq \frac{\mphi^2}{v}\,.
 \label{eq:phiphih_mass}
\end{equation}
Hence the coupling becomes independent of $\theta$. This observation is reflected in the limits on
$\left(\mphi, \sin\theta\right)$ derived from bounds on the $c_{\phi\phi h}$ coupling from
$h\to \phi\phi$ decays, see Sects.~\ref{sec:H_untagged} and \ref{sec:direct-exoH} as well as
Figs.~\ref{fig:indirectprobes} and \ref{fig:DirIndir}.

Furthermore, the mixing with the relaxion modifies the Higgs self-coupling $\lambda$ with respect to
its SM value of $\lambda_{\rm SM}=\frac{m_h^2}{2\,v^2}$ (at tree level). By demanding the heavier
mass eigenvalue to correctly reproduce the observed Higgs mass, $\lambda$ can be expressed as a
function of $f$, $\rbr$ and $s_0$ as
\begin{align}
 \lambda &= \frac{-f^2 m_h^4 + c_0\, m_h^2\, \rbr^4 v^4 + 4\rbr^8\, s_0^2\, v^6}{-2f^2 m_h^2 v^2 + 2c_0\, \rbr^4 v^6}
 \simeq \frac{f^2 - 4\rbr^4\,\left(c_0+16\rbr^4 s_0^2\right)\,v^2}{8\left(f^2 - 4c_0\,\rbr^4 v^2\right)}\,,
 \label{eq:lambda}
\end{align}
where the simplification holds for the approximation of $m_h\simeq v/2$. In addition, in order to
avoid a negative $\lambda$ and other overly large contributions to Higgs-couplings, we shall take
$ f>1\tev$ in our numerical analysis.

\subsection{Relaxion--gauge-boson interactions}
\label{sec:Rgauge}

An important ingredient for bounds on the relaxion parameter space are the couplings to the SM gauge
bosons.  The relaxion couples to the gauge bosons via two different classes of couplings, one that
arises through its mixing with the Higgs and one that can arise due to its pNGB nature and is
generically dictated by symmetry principles. The former couplings are found when \cp{} violation
(CPV) is present, as discussed above, while the latter are \cp{} conserving (CPC).

The leading terms in the effective Lagrangian that describes the \cp{}-even relaxion--gauge-boson
interactions are
\begin{align}
\mathcal{L} \supset -{\phi\, \sin\theta }\bigg(&
            \frac{\alpha}{16 \pi v}  F^{\mu\nu}F_{\mu\nu} + \frac{\alpha}{8\pi v \tan\theta_{\rm W}}  Z^{\mu\nu}F_{\mu\nu}
           + \frac{\alpha_s}{16  \pi v}  G^{\mu\nu,a}G^a_{\mu\nu}
           \nonumber
  \\
           &+ 2\frac{m_W^2}{v}  W^{+\,\mu}W^-_{\mu} 
           +\frac{m_Z^2}{v} Z^{\mu}Z_{\mu}\bigg) \,,
\label{eq:lagrangian}
\end{align}
where an order one factor that depends on $m_\phi$ is omitted for simplicity (see
\textit{e.g.}~\cite{Gunion:1989we}).  If some of the SM fields couple directly to the relaxion in a
manner that breaks the shift symmetry, then the above couplings will also contain non-universal
pieces that cannot be described only by $\sin\theta$, a fact that leads to model dependence.

As for the pseudoscalar couplings of the relaxion to the gauge fields, these arise if the
backreaction sector is anomalously charged under the corresponding gauge group, we thus write the
effective Lagrangian as
\begin{align}
  \mathcal{L} \supset
  \frac{\phi}{4\pi\,f}
  \left(
  \frac{\tilde c_{\gamma\gamma}}{4}
  F_{\mu\nu}
  \widetilde F^{\mu\nu}
  + \frac{\tilde c_{Z\gamma}}{2}Z_{\mu\nu} \widetilde F^{\mu\nu} + \frac{\tilde c_{ZZ}}{4}
  Z_{\mu\nu} \widetilde Z^{\mu\nu} + \frac{\tilde c_{WW}}{4} W_{\mu\nu} \widetilde W^{\mu\nu} +
  \frac{\tilde c_{GG}}{4}G_{\mu\nu} \widetilde G^{\mu\nu}\right)\,.
 \label{eq:LpsVV}
\end{align}
In order to compare the relevant importance of the pseudoscalar couplings to the ones induced by the
mixing with the Higgs let us consider for instance the couplings to photons.  For example, in cases
where the coupling to photons is induced via a weakly coupled set of fermion fields, we expect
$\tilde c_{\gamma\gamma}\sim \alpha Q_{\rm eff}^2$, where $Q_{\rm eff}$ stands for the chiral sum of
the fermion charges that induces the anomaly.  We note that if the backreaction sector is
anomaly-free, \textit{e.g.}~consists of a SM gauge-neutral sector (see for example the models
described in Refs.~\cite{Gupta:2015uea,Davidi:2018sii}), then $Q_{\rm eff}$ is suppressed by at
least one additional loop factor and is expected to be rather small~\cite{Flacke:2016szy}.

Two cases can be distinguished.
\begin{itemize}
\item There are two relevant couplings that are unique to the CPV sector: the triple scalar
  coupling, $c_{\phi\phi h} h\phi\phi\sim {\mphi^2}/{v}$, given in Eq.~\eqref{eq:phiphih_mass}; and
  the relaxion coupling to $WW$ ($ZZ$), defined in Eq.~\eqref{eq:lagrangian} as
  $c_{\phi WW\,(ZZ)}\sim s_\theta \,2\,m_W^2/v$ $(s_\theta \,2\,m_Z^2/v)$.  These couplings
  generically dominate the CPC ones at colliders, in particular when one can produce at least one
  electroweak/Higgs boson on shell. Thus play a crucial role below.
\item As for the coupling to photons, in this case both the CPV and CPC couplings are irrelevant in our effective description, and their ratio is given by 
\begin{equation} 
  \sin\theta \times  \frac{f}{v} \times \frac{1}{Q_{\textrm{eff}}^2}\sim   \frac{r_{\textrm{br}}^2}{Q_{\textrm{eff}}^2}\,.
\label{compariso}
 \end{equation}
 We find that the CPV coupling would dominate only for a relatively large backreaction scale, or in cases where the backreaction sector is non-anomalous.
 \end{itemize}

\section{Status of experimental probes for the relaxion}
\label{sec:stat-exper-prob}

As just summarized in Sect.~\ref{sec:RHmix}, the relaxion mixes with the SM-like Higgs, leading to a
scalar interaction between the relaxion $\phi$ and the SM particles. This provides the possibility
to search for such a particle in various experimental setups \cite{Flacke:2016szy,Choi:2016luu},
depending on its mass.  Its possible mass range spans several orders of magnitude from sub-eV to
several tens of GeV. Hence, it can be directly produced at high-energy colliders or be indirectly
detected through its long-range, spin-independent interaction between matter constituents.

\subsection{\texorpdfstring{\cp{}-even couplings}{CP-even couplings}}

\begin{figure}[tb]
  \centering
  \includegraphics[width=\textwidth]{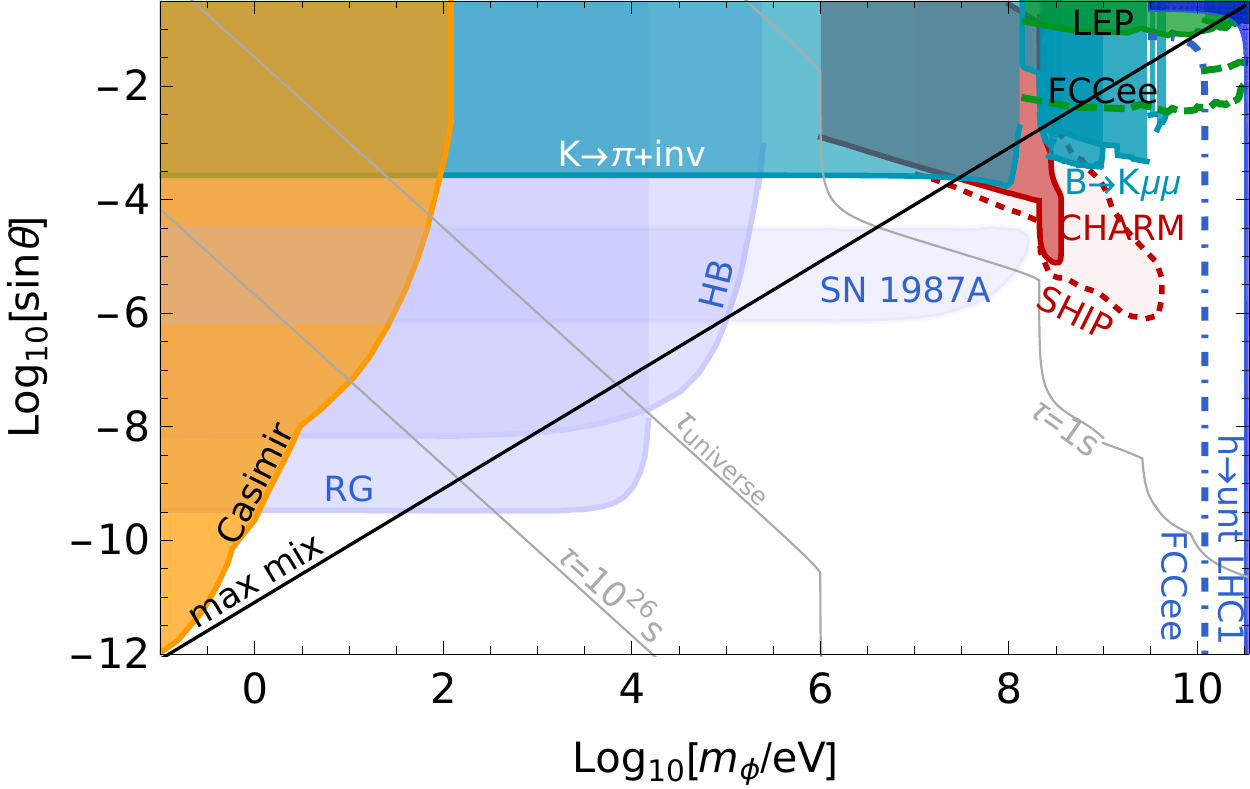}
  \caption{Summary of present bounds and few projections on the relaxion mass $\mphi$ and the mixing
    angle $\sin\theta$ (for details see Ref.~\cite{Flacke:2016szy}): Fifth force via the Casimir
    effect (orange)~\cite{Bordag:2001qi,bordag2009advances}, astrophysical probes (light
    blue)~\cite{Grifols:1986fc,Grifols:1988fv,Redondo:2013lna,Hardy:2016kme,SN1987A,Krnjaic:2015mbs}
    such as red giants (RG), horizontal branch stars (HB) and the Supernova (SN) 1987A, rare meson
    decays (turquoise) where the strongest bounds stem from $K\to\pi+\textrm{invisible}$ at
    E949~\cite{Artamonov:2009sz}, $K_L\to \pi l^+l^-$ at
    KTeV/E799~\cite{AlaviHarati:2000hs,AlaviHarati:2003mr} and $B\to K\mu^+\mu^-$ at
    LHCb~\cite{Aaij:2012vr,Aaij:2015tna}.  Beam dump experiment for $\phi$ production from $K$- and
    $B$-decays at CHARM~\cite{Bezrukov:2009yw,Clarke:2013aya,Schmidt-Hoberg:2013hba} and a
    projection from SHiP~\cite{Alekhin:2015byh} (red dotted).  Constraints from the $\phi Z$
    interaction (green) via $Z\to Z^*\phi$ and $e^+e^-\to Z\phi$ at
    LEP~\cite{Acciarri:1996um,Schael:2006cr} and projections for the same processes at the FCCee
    (green dashed). Untagged Higgs decays (blue) at the LHC Run-1~\cite{Flacke:2016szy} and
    projections for the FCCee and TeraZ (blue, dash-dotted, see Sect.~\ref{sec:H_untagged}).  The
    gray contours of the relaxion lifetime of $\tau_\phi=1$\,s, $10^{17}$\,s and $10^{26}$\,s
    indicate the beginning of BBN, the lifetime of the universe and safety from constraints of
    extragalactic background light, respectively.  The black line shows the upper bound on the
    mixing according to Eq.~(\ref{eq:maxmixNatural}).  }
  \label{fig:summary}
\end{figure}

Refs.~\cite{Flacke:2016szy,Choi:2016luu} studied the probes of the relaxion parameter space in
detail. We summarize their results in the overview plot in Fig.~\ref{fig:summary}.  The diagonal
line represents the maximal mixing as allowed by Eq.~\eqref{eq:maxmixNatural}.  One can classify the
following types of probes that are relevant for the parameter space below or near this line:
\begin{itemize}
\item \textbf{Fifth force experiments} for $m_{\phi} <10$\,eV. They comprise bounds from the Casimir
  effect~\cite{Bordag:2001qi,bordag2009advances} (shown here) and the E\"ot-Wash-type
  experiments~\cite{Smith:1999cr,Schlamminger:2007ht} for the equivalence
  principle~\cite{Spero:1980zz,Hoskins:1985tn,Chiaverini:2002cb,Hoyle:2004cw,Smullin:2005iv,Kapner:2006si}
  and the inverse square law~\cite{Hoyle:2004cw,Smullin:2005iv,Kapner:2006si} (for lower masses than
  shown here, see Ref.~\cite{Flacke:2016szy}).
\item \textbf{Astrophysical bounds} for $\mphi \lesssim 300\mev$. They stem from red giants,
  horizontal branch stars~\cite{Grifols:1986fc,Grifols:1988fv,Redondo:2013lna,Hardy:2016kme} and the
  Supernova 1987A~\cite{SN1987A,Krnjaic:2015mbs}.
\item \textbf{Beam dump experiments} for $1\mev<\mphi\lesssim 300\mev$. In particular proton fixed
  target experiments are sensitive. The presented bound is from
  CHARM~\cite{Bezrukov:2009yw,Clarke:2013aya,Schmidt-Hoberg:2013hba}.
\item \textbf{Rare meson decays} for $\mphi\lesssim 5\gev$. Relaxions can mediate rare decays of
  $K,B$ and $\Upsilon$ mesons, hence their branching ratios constrain the relaxion-Higgs mixing
  angle, see \textit{e.g.}~Refs.~\cite{Flacke:2016szy,Artamonov:2009sz,Aaij:2012vr,Aaij:2015tna} and
  references therein.
\item \textbf{LEP} for $0.3\gev<\mphi<116\gev$. Searches for Higgs-like particles produced in the
  $Z$ decay $Z\to f\bar f \phi$ and via Higgs-strahlung $e^+ e^- \to \phi Z$ have been
  performed~\cite{Acciarri:1996um,Schael:2006cr}.
\item \textbf{Higgs decays} for $\mphi<m_h/2$. The LHC Run-1 set an upper limit on the untagged
  branching ratio of the Higgs boson~\cite{Flacke:2016szy,Bechtle:2014ewa}.
\end{itemize}
In addition, the relaxion is constrained by cosmological bounds on its late decays.  Significant
regions of the parameter space can be probed by big bang nucleosynthesis (BBN), the cosmic microwave
background (CMB) and extragalactic background light (EBL) as discussed in
Refs.~\cite{Flacke:2016szy, Fradette:2017sdd}.  We do not present these bounds here because they
depend on various details like the reheating temperature, resulting in a strong model dependence
that may even allow to circumvent them~\cite{Choi:2016kke}.  Instead, in Fig.~\ref{fig:summary} we
show only lifetime contours to highlight where cosmological probes could play an important role, in
particular 1\,s as an indication of the beginning of BBN, $ 10^{17}$\,s as the lifetime of the
universe, and $10^{26}$\,s to indicate where the relaxion is unbounded by EBL~\cite{redondo}.

In the 10\,eV--1\,MeV region, only astrophysical probes are sensitive enough to probe relevant
regions, \textit{i.e.}~below the line from Eq.~\eqref{eq:maxmixNatural}.  Together with the bounds
from invisible kaon decays shown in Fig.~\ref{fig:summary}, neutron scattering experiments and
atomic precision measurements can in principle probe Higgs portal models, but they are not yet
sensitive enough to set competitive bounds~\cite{Pokotilovski:2006up, Nesvizhevsky:2007by,
  Frugiuele:2016rii, Berengut:2017zuo, Delaunay:2017dku, Arvanitaki:2014faa, Hees:2018fpg,
  Arvanitaki:2017nhi, Delaunay:2016brc}. Above 1\,MeV, rare meson decays, beam dump and collider
experiments have sensitivity to the physical parameter region below the line.

Future experiments are being planned to improve on these existing limits.  For instance both
SHiP~\cite{Alekhin:2015byh} and the beam dump run of NA62~\cite{Dobrich:2015jyk} will extend the
coverage in the MeV--GeV mass range.  Moreover, the MATHUSLA surface detector will be able to
constrain the couplings of light, long-lived scalars below $\mphi\simeq 5\gev$~\cite{Evans:2017lvd,
  Curtin:2018mvb}. In the following we will analyse how a relaxion in the 5\,GeV--35\,GeV mass range
can be probed by the HL-LHC and future lepton colliders.  Heavier relaxions with a mass of
35\,GeV--62.5\,GeV are already ruled out by the present limits on untagged Higgs decays as shown in
Ref.~\cite{Flacke:2016szy}.  Other scalars that mix with the Higgs could still exist in this mass
range.

\subsection{\texorpdfstring{\cp{}-odd couplings}{CP-odd couplings}}
\label{sec:cp-odd-couplings}

Pseudoscalar couplings of the relaxion originate from the backreaction sector, and thus they depend
on the details of the specific model as we briefly discussed in Sect.~\ref{sec:RHmix}.  The
perspectives to probe the relaxion via these $\cp$-odd couplings are subject to the relative size of
the $\cp$-odd and -even couplings.
\par
In the following we focus on the mass region above 5\,GeV and we refer to
Refs.~\cite{Knapen:2016moh, Jaeckel:2015jla, Bauer:2017ris} for a detailed discussion of the
phenomenology of axion-like particles with $\sin{\theta} =0$.  As it was shown in
Ref.~\cite{Knapen:2016moh}, heavy-ion collisions at the LHC can provide the best limits on $\cp$-odd
photon couplings in the 5\,GeV--100\,GeV mass range. Considering Pb-Pb collisions with a luminosity
of 1\,nb$^{-1}$ yields
\begin{equation}
\frac{f}{ \tilde{c}_{\gamma \gamma}} \gtrsim 500 \gev\,.
\end{equation} 
LEP mono-photon searches provide a slightly weaker bound,
$f/\tilde{c}_{\gamma\gamma} \gtrsim 300 \gev$~\cite{Jaeckel:2015jla}.

The $\phi\gamma Z$ dual coupling is instead constrained by rare $Z$ decays. For Higgs-like decays of
the relaxion, the strongest limit of
\begin{equation}
\frac{f}{ \tilde{c}_{Z \gamma}}  \gtrsim 1  \;  \rm TeV
\end{equation}
comes from ALEPH~\cite{Barate:346643,ALEPH:2005ab} as we derive in
Sect.~\ref{sec:probing-cp-odd}. This decay pattern does not only appear when the Higgs-relaxion
mixing dominates, but also in scenarios with $\cp$-odd couplings to SM fermions as in
Ref.~\cite{Davidi:2017gir}. The case of \mbox{$BR( \phi \to \gamma \gamma )=1 $} was studied in
Ref.~\cite{Bauer:2017ris}.
Similarly, the analogous $\phi Z Z$ coupling gives rise to the rare $Z$ decay
$ Z \to \phi Z^*(f \bar f)$, and to a weak bound via the additional contribution to the total
$Z$-width, $\Gamma_Z^\text{NP}$~\cite{Bauer:2017ris}.

Furthermore, since the relaxion has both $\cp$-odd and -even couplings, they possibly generate a
contribution to electric dipole moments (EDMs). The strongest bound comes from the electron EDM,
$d_e$. In particular the present upper bound of
$ d_e/e \sim 8 \times 10^{-29}\,\text{cm}$~\cite{Baron:2013eja} implies for the product of the
$\cp$-odd coupling to photons and $\cp$-even one to electrons~\cite{Flacke:2016szy},
\begin{eqnarray}
  \frac{ \alpha \tilde{c}_{\gamma\gamma}}{ \pi} \frac{m_e}{v} \frac{ \sin{\theta}}{f} \lesssim  2
  \times  10^{-13}\gev^{-1} ~\Rightarrow~ \frac{f}{\tilde{c}_{\gamma \gamma}} >  2.5\times 10^4 \sin{\theta}\gev\,,
  \label{eq:edm}
\end{eqnarray} 
where we omitted the logarithmic mass dependence (and, for simplicity, we omit the model dependent
\cp{}-odd couplings to fermions here).  In the coming years improvements of one order of magnitude
are expected \cite{Hewett:2012ns}.  Consequently, as we will see in the following, a relaxion signal
could be expected at future EDM experiments and high energy colliders, such as the HL-LHC or future
electron colliders.

\section{Prospects for  colliders}
\label{sec:prospecs-at-high}

After reviewing the constraints on the relaxion mass and mixing angle in the previous section, we
will now identify and discuss different sensitive channels at hadron and electron colliders that are
able to improve on the existing limits. The region of relaxion masses between 5\gev{} and 35\gev{}
is so far only poorly constrained, but can be probed in various ways by runs with higher luminosity
at the LHC as well as at future colliders. Higgs searches at LEP and constraints from untagged Higgs
decays at the LHC set the so-far strongest limits in this region.

Before going to the details of the collider channels, we notice that both very low $f\ll M_{\rm Pl}$
and $\rbr\lesssim \mathcal{O}(1)$ are required in order to obtain a relaxion mass in the GeV
range~\cite{Flacke:2016szy,Choi:2016luu}, which implies that only the little hierarchy can be
addressed by the relaxion.  For $\mphi>5\gev$, Eq.~\eqref{eq:exact} implies the following relations
between the underlying model parameters:
\begin{align}
  f \geq \left\lbrace 1\tev,~10\tev  \right\rbrace  &\implies \rbr \gtrsim  \left\lbrace  0.3,~0.9  \right\rbrace \\
  s_0 \geq \left\lbrace 1/\sqrt{2} ,~0.1,~0.01 \right\rbrace &\implies f \lesssim \left\lbrace 1.5\tev,~15\tev,~150\tev \right\rbrace \\
  \rbr \leq \left\lbrace 1,~1.5 \right\rbrace &\implies f \lesssim \left\lbrace 10\tev,~30\tev \right\rbrace \,.
\end{align}
With $f$ and $\Lambda_{\rm br}\equiv r_{\rm br} v$ potentially within the range of future colliders,
there is a chance to encounter the richer phenomenology of the UV completion of a relaxion model,
beyond a discovery of the relaxion itself.  Moreover, the heavier the relaxion, the looser the upper
bound on the mixing angle $\sin\theta$, see Eq.~\eqref{eq:maxmixNatural}.

In the following we will use the relaxion as a benchmark model. However, our results are more
universal and apply also to a general Higgs portal.  The only difference is the model-dependence of
the quartic Higgs coupling $\lambda$ given in Eq.~\eqref{eq:lambda} and the triple-scalar coupling
$c_{h\phi\phi}$ given in Eq.~\eqref{eq:phiphih}.  Hence our interpretation of the limit on the
$h\to\phi\phi$ decay is relaxion-specific, while all the other presented bounds are general.

\subsection{Precision probes}
\label{sec:indirect-probes}
In this section we study possible constraints on the relaxion parameter space via precision
measurements of Higgs and $Z$ properties, both at the HL-LHC and future lepton colliders.

\subsubsection{Untagged Higgs decays}
\label{sec:H_untagged}

A powerful way to probe the relaxion at colliders is the exotic Higgs decay into a pair of
relaxions, $h\to \phi\phi$~\cite{Flacke:2016szy}, exploiting the sensitivity to the triple coupling
$c_{\phi\phi h}$ given in Eq.~(\ref{eq:phiphih}). There are two complementary ways of deriving
bounds from this decay. One possibility are direct searches for the decay products of the relaxions;
the other one is indirect by virtue of the NP contribution to the Higgs width in a global fit of
Higgs couplings.  In this Section we focus on the latter constraint, whereas we review the various
direct searches in Sect.~\ref{sec:direct-exoH}.  In Fig.~\ref{fig:DirIndir} both approaches are
compared, showing their strong potential in particular at lepton colliders.

Global Higgs coupling fits are sensitive to additional exotic decay channels for the Higgs that
remain untagged in the corresponding searches.  The measured Higgs rates allow for a global fit of
the Higgs coupling modifiers $\kappa_i$ and the branching ratio into NP $\br(h\to \np)$ as an
additional parameter.  Because the measured Higgs rates consist of the product of production and
decay, the fit is model dependent.  For the case of the relaxion mixed with the Higgs, a
two-parameter fit is required, namely a universal modifier of the Higgs couplings to SM particles
that can be identified as $\kappa\equiv \cos\theta$ (thus automatically $\kappa_V\leq 1,~V=W,Z$),
and $\br(h\to \np)$ that is realized by the $h\to\phi\phi$ decay channel.  The total Higgs width is
given by
\begin{equation}
  \Gamma_h = \kappa^2\, \Gamma_h^{\rm SM} + \Gamma_h^{\np}\,.
  \label{eq:Ghtot}
\end{equation}
In general, the NP contribution to the Higgs width consists of
\begin{equation}
  \Gamma_h^{\np} = \Gamma_h^{\inv} + \Gamma_h^{\unt}\,,
\end{equation}
where $\Gamma_h^{\inv}$ denotes the partial width into \textit{invisible} particles and
$\Gamma_h^{\unt}$ denotes the partial width into \textit{untagged} final states that are not
necessarily undetectable, but were not accounted for in the data set included in the fit, see
\textit{e.g.}~Refs.~\cite{Heinemeyer:2013tqa, Khachatryan:2014jba, Bechtle:2014ewa,
  CMS-PAS-HIG-17-031, ATLAS-CONF-2018-031}.  In the relaxion case, we are interested in constraining
$\Gamma_h^{\np}=\Gamma_h^{\unt}=\Gamma(h\to\phi\phi)$, \textit{i.e.}~not the invisible width.  With
masses in the GeV range, the relaxion is short-lived and decays inside the detector even for small
$\sin\theta$~\cite{Flacke:2016szy}. For the case of only one universal coupling modifier $\kappa$,
Ref.~\cite{Bechtle:2014ewa} obtains ${\rm BR}(h\to {\rm NP})\leq 20\%$ from Run-1 data of the LHC
and Ref.~\cite{Belanger:2013xza} reports a similar bound. This bound applies directly to the
relaxion setup, as long as the modifications of the Higgs phenomenology are sufficiently described
by $\sin\theta$ and the impact of the backreaction sector remains small.

Current and future runs at the \mbox{(HL-)}LHC as well as Higgs precision measurements at future
lepton colliders will tighten the bound on untagged Higgs decays.  In Tab.~\ref{tab:coll_BRuntagged}
we collect the projections and estimates of upper bounds on ${\rm BR}(h\to{\rm untagged})$ at the
\mbox{(HL-)}LHC, ILC, CEPC, CLIC and FCCee running at different energies and luminosities, and
compare them to the bounds on ${\rm BR}(h\to{\rm invisible})$ from the literature.

\begin{table}[tb]
 \begin{center}
  \begin{tabular}{|l|c|r||l|c||l|c|}
   \hline
   Collider &$\sqrt{s}$ [TeV]&$\lint$ [fb$^{-1}$]& $\textrm{BR}_{\rm inv}$ [\%]& Ref. &$\textrm{BR}_{\np}$ [\%]&Ref.\\ \hline\hline
   LHC1    &$7, 8$		&22	&37	&\cite{Bechtle:2014ewa,Belanger:2013xza}&20&\multirow{3}{*}{\cite{Bechtle:2014ewa}}\\
   LHC3	    &$13$	&300	&8.8 (68\%)	&\cite{Bechtle:2014ewa}	&7.6 (68\%)&\\
   HL-LHC   &$13$	&3\,000	&5.1 (68\%)	&\cite{Bechtle:2014ewa}	&4.3 (68\%)&\\ \hline \hline
   CLIC	    &0.38	&500	&0.97 (90\%)&\cite{Abramowicz:2016zbo,CLIC:2016zwp}&3.1&est.\\ \hline
   CEPC	    &0.25	&5\,000	&1.2&\cite{Chen:2016zpw}		&1.9&est.	\\ \hline
   ILC      &0.25	&2\,000	&0.3	&\cite{Fujii:2017vwa}&1.5 &est.\\\hline
   FCCee    &0.24	&10\,000	&0.19	&\cite{Dawson:2013bba,Gomez-Ceballos:2013zzn}&0.64 &est.\\ \hline
  \end{tabular}
  \caption{Current upper bound and projections on the branching ratios of $h\to \textrm{invisible}$
    and $h\to\np$ at various colliders running at the given center-of-mass energies $\sqrt{s}$ for
    benchmarks of integrated luminosity $\lint$. Unless states otherwise in parenthesis, the bounds
    are given at the 95\% CL. The $\br(h\to\np)$ at future lepton colliders is estimated via
    Eq.~(\ref{eq:BRdeltakappa}) and the precision on $\kappa_Z$ given in the text. }
  \label{tab:coll_BRuntagged}
 \end{center}
\end{table}

\paragraph{Estimate of $\br\left(h\to\np\right)$}
While projections of the sensitivity to invisible Higgs decays have been published for all
considered lepton colliders~\cite{Dawson:2013bba,Gomez-Ceballos:2013zzn, Bechtle:2014ewa,
  Chen:2016zpw, Abramowicz:2016zbo, CLIC:2016zwp, Fujii:2017vwa}, to our knowledge such systematic
studies are still missing for Higgs decays to untagged final states. Therefore we estimate the upper
bounds on $\br(h\to\np)$ in the following way.

Due to the NP contribution to the total Higgs width in Eq.~(\ref{eq:Ghtot}), each branching ratio
into SM final states $F$ is diluted as
\begin{equation}
 \br(h\to F) = \br^{\sm}(h\to F)\,\cdot\,\left[1-\br(h\to \np)\right]\,.
 \label{eq:BRdilution}
\end{equation}

A bound can be set on $\br(h\to\np)$ via the precision $\delta_\kappa$ of the experimental
determination of $\kappa$~\cite{Bechtle:2014ewa}\footnote{We thank Tim Stefaniak for a helpful
  discussion regarding the approximate bounds on untagged Higgs decays at lepton
  colliders.}. Assuming that the observed signal rates of production times decay are SM-like within
$n$ times the uncertainty, $n\cdot \delta_\kappa$, the modification of the production and decay must
result in the product as
\begin{equation}
 \left(1 - n\cdot \delta_\kappa\right)^2 \leq \kappa^2\,\cdot\,\left[1-\br(h\to \np)\right]\,.
 \label{eq:SMlike}
\end{equation}
Hence, the branching ratio into NP can be bounded as
\begin{equation}
 \br(h\to\np) \leq 1 - \left(\frac{1-n\cdot \delta_\kappa}{\kappa} \right)^2\,.
 \label{eq:BRdeltakappa}
\end{equation}

Coupling fits for a model with one universal modifier have not been performed for all of the
colliders considered here. Instead, several coupling modifiers for different particle species have
usually been included. However, the by far most precisely determined coupling at lepton colliders is
generally $\kappa_Z$, \textit{i.e.}~the coupling of $hZZ$ due to the absolute measurement of the
$Zh$ associated production cross section. Therefore this parameter dominates in the fit and is
expected to yield a similar result as in a one-parameter fit. A fit of only one $\kappa$ and
$\br(h\to\np)$ would result in a stronger bound; thus the described approximation can be regarded as
a conservative estimate.

In the relaxion case, $\br(h\to\np)\equiv \br(h\to\phi\phi)$ depends on $\mphi$ and
$\kappa\equiv\sqrt{1-\sin^2\theta}$. Thereby, as shown in Fig.~\ref{fig:indirectprobes}, we set
constraints on $(\mphi,\sin^2\theta)$ approximately at the $95\%$~CL. We obtain these by setting
$n=2$ in Eq.~(\ref{eq:BRdeltakappa}), and using the projected precision of $\kappa_Z$ from the
various lepton colliders: $\dkaZ^{\rm CLIC} = 0.8\%$~\cite{Abramowicz:2016zbo},
$\dkaZ^{\rm CEPC} = 0.49\%$~\cite{Chen:2016zpw}, $\dkaZ^{\rm ILC} = 0.38\%$~\cite{Fujii:2017vwa},
$\dkaZ^{\rm FCCee} = 0.16\%$~\cite{Dawson:2013bba}.

In the limit $\kappa=1$, a conservative bound on $\br(h\to\np)$ can be derived.  We summarize these
bounds for future lepton colliders in Tab.~\ref{tab:coll_BRuntagged} (denoted by "est") derived from
the precision goals of $\kappa_Z$.  The resulting exclusion contours based on these
$\kappa$-independent BR bounds are almost identical to those from constraining the product of
$\kappa$ and the BR in Eq.~(\ref{eq:SMlike}) because $\kappa$ is very close to 1 for the
$\sin^2\theta$ of interest and the $\kappa$-dependence of the product is dominated by the
$\kappa$-dependence of $\br(h\to\phi\phi)$.

For the FCCee, the Higgs BR into NP has been fitted in a multi-$\kappa$ fit without requiring
$\kappa_V\leq 1$, resulting in $\br(h \to\np) \leq 0.48\%$ at the $1\sigma$
level~\cite{Gomez-Ceballos:2013zzn}. The order of magnitude agrees well with our $2\sigma$ estimate
of $0.64\%$, but due to these two restrictions we apply our result for a coherent comparability
among the lepton colliders.

For the LHC, $\br(h\to\np)$ has been worked out in a global Higgs coupling fit for different
luminosities in Ref.~\cite{Bechtle:2014ewa}. The bound based on the Run-1 data set is explicitly
calculated in the model with one universal coupling modifier and a NP branching ratio, and is also
presented at the 95\%~CL. In contrast, the projections for 300 and 3000$\ifb$ result from a
multi-dimensional $\kappa$ fit and are reported only at the $68\%$~CL. The two modifications with
respect to the Run-1 bound have opposite effects and partially compensate each other. A dedicated
global fit of Higgs couplings to the single-$\kappa$ case is necessary in order to obtain $95\%$ CL
bounds precisely applicable to the relaxion case.

\begin{figure}[ht!]
 \begin{center}
  \includegraphics[width=0.6\textwidth]{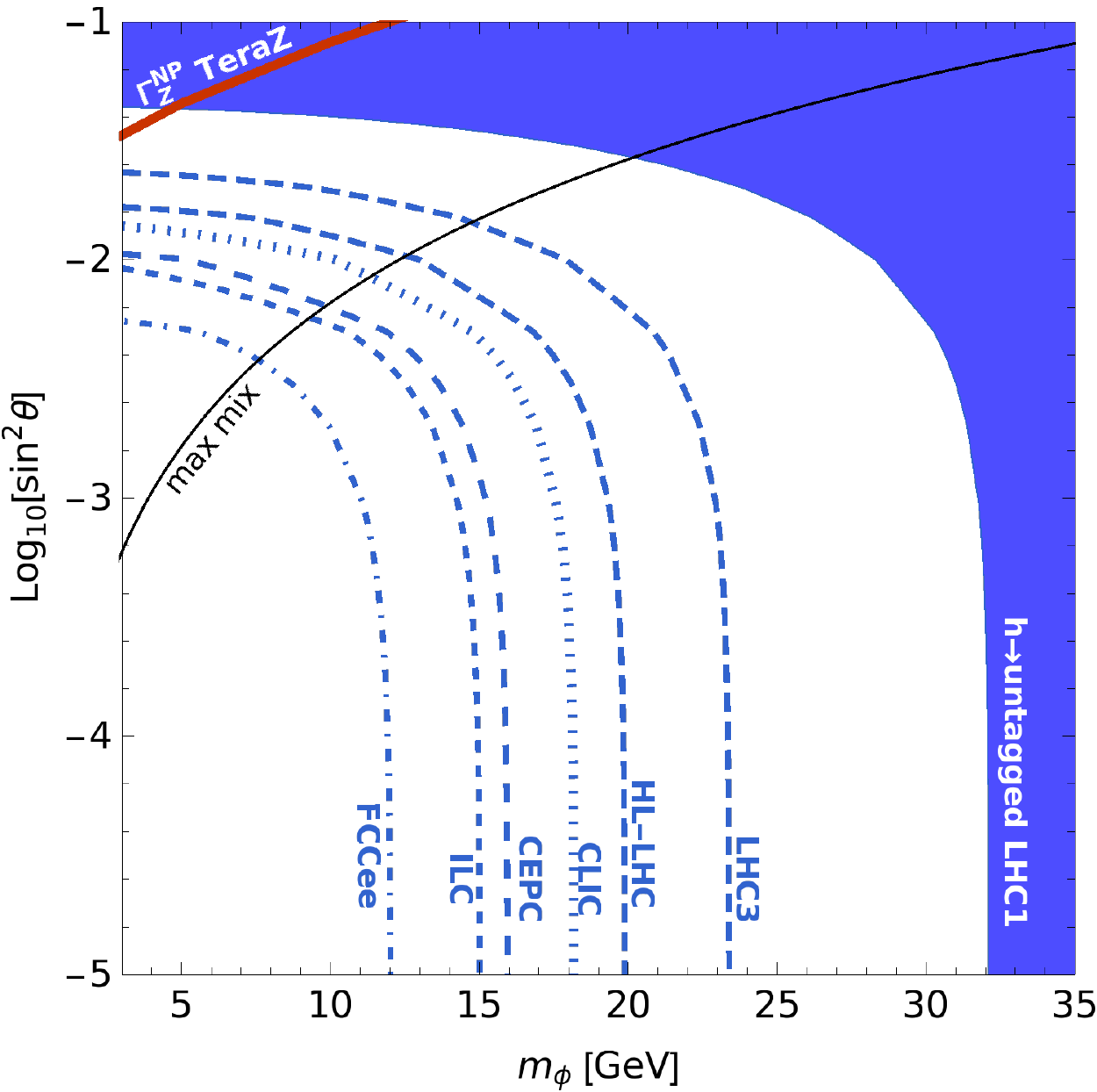}
  \caption{\emph{Precision} bounds on $\st$ and $\mphi$: Upper limit on the untagged branching ratio
    of the Higgs boson, here $h\rightarrow \phi\phi$ (blue), obtained via the precision of Higgs
    couplings.  Current (solid, blue area) and projected (blue, dashed) exclusion from the (HL-)LHC,
    CLIC at $\sqrt{s}=380\gev$ with $500\ifb$, CEPC at $250\gev$ with $5\iab$, ILC with $250\gev$
    with $2\iab$ and FCCee at $240\gev$ with $10\iab$.  The energies, luminosities and upper bounds
    on $\br(h\to\np)$ of the collider benchmarks are summarized in Tab.~\ref{tab:coll_BRuntagged}.
    The contours represent the 95\% CL, except for LHC3 and HL-LHC which are at 68\% CL.  Projection
    of the constraint on the NP contribution $\Gamma(Z\to \phi f\bar f)$ to the total $Z$-width,
    assuming the experimental precision of the FCCee running at the $Z$-pole with $10^{12}Z$ and an
    improved theory uncertainty (red).  The black line shows the upper bound on the mixing according
    to Eq.~(\ref{eq:maxmixNatural}).}
  \label{fig:indirectprobes}
 \end{center}
\end{figure}

\paragraph{Implication of the untagged Higgs decays on the relaxion parameter space}
The fact that the coupling does not vanish in the limit of small $\sin\theta$---owing to the
$\cos^3\theta$ term in Eq.~(\ref{eq:phiphih_mass})---gives rise to a $\sin\theta$-independent bound
on the mass for small $\sin^2\theta \lesssim 10^{-3}$.  Fig.~\ref{fig:indirectprobes} shows in blue
the projections arising from the ${\rm BR}(h\to{\rm untagged})$ bounds at future colliders. While
the present bound of the LHC Run 1 excludes $\mphi\gtrsim 32\gev$, the HL-LHC is expected to exclude
$\mphi\gtrsim 20\gev$. These bounds may be significantly improved by future lepton colliders. In
particular, the potential of the FCCee with $10\iab$ allows for an exclusion of
$\mphi\gtrsim 12\gev$. Since the $h\phi\phi$ coupling is $s_0$-dependent, also the exact exclusion
contours depend on the endpoint of the rolling. However, by comparing numerically $s_0$ of
$\mathcal{O}(1)$ and $\mathcal{O}(10^{-2})$, we notice that the impact on the
${\rm BR}(h\rightarrow \phi\phi)$ is only very mild. Consequently, these strong bounds can be
regarded as quite robust within the relaxion framework.

In contrast, for general Higgs portal models, the $h\phi\phi$ coupling has a different dependence on
the mixing angle and accordingly, the region of $\mphi>35\gev$ is not necessarily ruled
out. Consequently, searches for light scalars with such masses remain highly relevant.

\subsubsection{Higgs self-coupling}
\label{sec:higgs-quart-coupl}
The knowledge of the the Higgs self-coupling $\lambda$ is crucial for the understanding of the Higgs
potential.  As shown in Eq.~(\ref{eq:lambda}), the Higgs-relaxion mixing alters $\lambda$. However,
the sensitivity of present and future colliders to $\lambda$ via Higgs pair production is
limited. While relative deviations $\Delta\lambda/\lambda = \lambda/\lambda_{\rm SM}-1$ from the SM
value of 50\% are estimated to be in the reach of the HL-LHC~\cite{slides:CLIC2016}, the ILC might
reach 10\%~\cite{slides:CLIC2016}\footnote{However, this estimate of 10\% at the ILC is not
  presented with a detailed quantitative analysis.} whereas Ref.~\cite{DiVita:2017vrr} obtains -34\%
to +42\%.  According to Ref.~\cite{Abramowicz:2016zbo}, CLIC is expected to become sensitive to a
deviation of $\Delta \lambda/\lambda=19\%$ for an electron polarisation of $P^-=-80\%$ and combining
runs at $1.4\tev$ and $3\tev$ with luminosities of $1.5\iab$ and $2\iab$, respectively, whereas
Ref.~\cite{DiVita:2017vrr} reports -18\% to +28\% at the 68\% CL.  Combining the anticipated
measurements at the HL-LHC and the FCCee in a 13-parameter fit, Ref.~\cite{DiVita:2017vrr} concludes
a sensitivity of 40\%.

While the estimates of the different references may not be directly comparable, it becomes clear
that relatively sizeable deviations in the order of several $10\%$ of the Higgs self-coupling from
its SM prediction will remain in accordance with uncertainties at future colliders. Yet, the
relaxion with a mixing of $\sin^2\theta<0.1$ induces deviations from $\lambda_{\rm SM}$ of only less
than 10\%. Consequently, the considered future colliders will not be able to constrain the relaxion
parameter space via this indirect probe.  A precision of $1\%$ ($0.1\%$) would be needed to
constrain $\sin^2\theta<10^{-2}$ ($10^{-3}$).

\subsubsection{\texorpdfstring{Total $Z$ width}{Total Z width}}

Relaxion-mediated decays of the $Z$ boson give rise to an additional contribution
$\Gamma_Z^{\rm NP}$ to the total $Z$ width. Since the total width measurement at LEP1 is in
agreement with the theoretical SM prediction~\cite{ALEPH:2005ab,Patrignani:2016xqp}, a bound on the
NP contribution can be derived by limiting it to the combined experimental and theoretical
uncertainty. Currently the uncertainty is dominated by the experimental one of
$\delta\Gamma_Z^{\textrm{LEP1}}=2.3\mev$~\cite{ALEPH:2005ab} while the theoretical one is
$\delta\Gamma_Z^\textrm{th}=0.5\mev$~\cite{Freitas:2014hra}. At TeraZ, the experimental uncertainty
is expected to shrink below $\delta\Gamma_Z^\textrm{TeraZ} = 0.1\mev$~\cite{Gomez-Ceballos:2013zzn},
necessitating further theoretical improvement. When including the so far missing 3-loop
contributions, the theoretical uncertainty is estimated to reach
$\delta\Gamma_Z^\textrm{th,3-loop}=0.2\mev$~\cite{Freitas:2016sty}. Hence, unless the improvement of
the theory uncertainty goes beyond this estimate, the theoretical and experimental uncertainties
will be comparable, and we will use the combined projection of
$\Gamma_Z^{\rm NP}< 2\times\delta \Gamma_Z^{\rm th+exp}\simeq 2\times 0.22\mev$ to set a bound at
the 95\% CL. Saturating this bound by the relaxion-mediated 3-body decay $Z\to \phi f\bar f$ for
vanishing \cp{}-odd couplings, we obtain the constraint shown as a red line in
Figs.~\ref{fig:indirectprobes} and \ref{fig:DirIndir}.

This partial width can be expressed in terms of the \cp{}-even and -odd couplings of the relaxion,
giving rise to the integrands $\mathcal{I},~\widetilde{\mathcal{I}}$, respectively,
\begin{align}
  \label{eq:partial-width}
  \Gamma_Z^{\rm NP} &=
  \Gamma(Z\to \phi f\bar f)
  = \frac{1}{m_Z^3 \left(2\pi\right)^3}\sum\limits_{f}N_f^c \int\limits_{4m_f^2}^{(m_Z-m_\phi)^2}
           \textrm{d}m_{12}^2 \left(\mathcal{I} + \widetilde{\mathcal{I}}\right)\,.
\end{align}
The sum is over the fermion final states $f$ of interest, and $m_f$ and $N_f^c$ are the mass and
number of colors of the respective fermion. We obtain the following \cp{}-even and \cp{}-odd
integrands
\begin{align}
  \label{eq:cp-even-integrand}
  \begin{split}
  \mathcal{I} &= \frac{\sint^2}{18} \,\frac{\alpha\, \pi}{\cwquart \swquart} \,\sqrt{1
                -\frac{4m_f^2}{m_{12}^2}}\,\frac{\sqrt{\lambda_{\rm K}(m_Z, m_\phi,
                m_{12})}}{\left(m_{12}^2-m_Z^2\right)^2
                +m_Z^2\Gamma_Z^2}\times\\
  &\quad\left[\left(m_{12}^2+2m_f^2\right)\left(\frac{\lambda_{\rm K}(m_Z,m_\phi,m_{12})}{m_{12}^2}+12 
    m_Z^2\right)\left(a_f^2 + v_f^2\right) -72\, m_f^2\, \mZsq \,a_f^2\right]\,,
    \end{split}\\
\begin{split}
  \widetilde{\mathcal{I}} &=\frac{\alpha \,\pi}{36\left(4\pi f\right)^2}
                            \,\sqrt{1-\frac{4m_f^2}{m_{12}^2}}
                            \,\lambda_{\rm K}^{3/2}(\mZ,\mphi,m_{12})
                            \left[\tilde c_{\gamma Z}^2 Q_f^2 \frac{m_{12}^2+2m_f^2}{m_{12}^4} \right.
  \\
              &\quad \left. +\frac{\tilde c_{ZZ}^2}{\swsq \cwsq}\frac{\left(m_{12}^2 +
                2m_f^2\right)\left(a_f^2+v_f^2\right) - 6 m_f^2 a_f^2}{\left(m_{12}^2-\mZsq\right)^2+\mZsq
                \Gamma_Z^2}\right.\\
              &\quad\left.-\frac{2\,\tilde c_{\gamma Z}\,\tilde c_{ZZ} \,Q_f\, v_f}{\sw\cw}\,\frac{\left(m_{12}^2 +
                2m_f^2\right)\left(\mZsq-m_{12}^2\right)}{m_{12}^2\left[\left(m_{12}^2-\mZsq\right)^2
                + \mZsq \Gamma_Z^2\right]}\right]\,,
                \end{split}
\end{align}
respectively, with $a_f$, $v_f$ being the axial- and vector-coupling of the fermion, $Q$ its
electric charge, $\theta_\text{W}$ the Weinberg angle, and the K\"all\'en function is given by 
\begin{equation}
  \label{eq:mass-difference}
  \lambda_{\rm K}(m_1,m_2,m_3) = -(m_1+m_2+m_3)(m_1+m_2-m_3)(m_1-m_2+m_3)(-m_1+m_2+m_3)\,.
\end{equation}
In the calculation of $\mathcal{I}$ we neglected the effect of the loop-induced $\phi Z\gamma$
interaction. We find agreement of the \cp{}-even contribution and the result reported in
Ref.~\cite{Gunion:1989we}. For $\mphi=0$ and summing over all kinematically allowed SM fermions we
obtain
\begin{align}\frac{\Gamma}{\textrm{MeV}} &\approx 23 \, \sin^2 \theta + 2.1\times10^{-5}
                                           \left(\frac{\textrm{TeV}}{f}\right)^2 \left(1.1\,\tilde
                                           c_{ZZ}^2 - \tilde c_{ZZ}\tilde c_{\gamma Z} + 47\,\tilde
                                           c_{\gamma Z}^2\right)\,,
\end{align}
where the approximation is valid for $\mphi \lesssim 0.5\gev$ for the \cp{}-even and
$\mphi \lesssim 10\gev$ for the \cp{}-odd contribution. However, the latter contribution is
completely negligible in the range of interest. This justifies showing only the $\sin^2\theta$
contribution in Figs.~\ref{fig:indirectprobes} and \ref{fig:DirIndir}. In principle, also other
electroweak precision observables could play a role.  Yet, Ref.~\cite{Falkowski:2015iwa} showed that
for the considered masses they do not constrain small enough mixing angles in the Higgs portal.

\subsection{Direct probes}
\label{sec:direct-probes}
In a complementary way to the indirect bounds discussed in the previous section, the relaxion
parameter space can also be constrained by direct searches in various production modes as discussed
below.

\subsubsection{Pair production in Higgs decay}
\label{sec:direct-exoH}

While Higgs coupling fits are sensitive to the BR of $h\to\phi\phi$ irrespective of the decays of
the relaxions (see Sect.~\ref{sec:H_untagged}), one can also look directly for the relaxion decay
products.  Each relaxion, pair-produced in the Higgs decay, further decays into a pair of fermions
$f$, photons $\gamma$ or gluons $g$ resulting in a four-particle final state $F$. ATLAS and CMS
search for such signatures and report $\mphi$-dependent bounds on
$\frac{\sigma_h}{\sigma_h^{\rm SM}} \times {\rm BR}\left(h\to\phi\phi \to F\right)$, which can be
compared to the prediction in the relaxion framework and thereby be translated into a bound on
$\sin\theta$ and $\mphi$.  In Tab.~\ref{tab:exoH} we summarize the status of these exotic Higgs
decay searches performed by ATLAS and CMS during Run 1 and Run 2 of the LHC, listing the final
states, the considered data set and mass range of $\mphi$. We conclude that none of the current
searches is sensitive enough to probe parts of the relaxion parameter space displayed in
Figs.~\ref{fig:indirectprobes} and \ref{fig:DirIndir}, \textit{i.e.}~$5\gev \leq \mphi \leq 35\gev$
and $10^{-5}\leq \sin^2\theta \leq 10^{-1}$.  Moreover, we estimate the potential reach of the
HL-LHC with $3000\ifb$ by rescaling the current limits by the ratio of luminosities and, in the case
of Run 1 limits at $\sqrt{s}=8\tev$, additionally by the ratio of Higgs production cross sections at
$8$ and $13\tev$ in the dominant channels~\cite{xsecratio}. In Tab.~\ref{tab:exoH} we state the
relaxion mass corresponding to the vertical asymptote of the exclusion contour in the
$(\mphi, \sin^2\theta)$ plane.

The strongest direct bound at the HL-LHC is expected in the $bb\tau\tau$ channel excluding $\mphi>26\gev$ which is presented
in Fig.~\ref{fig:DirIndir} (orange, dashed). The projections for $bb\mu\mu$ and $4b$ with
$Vh, V=W,Z$ production are similar, but somewhat weaker (see $\mphi^{\rm HL}$ in Tab.~\ref{tab:exoH}) and therefore not
shown in the overview plot.  Neither the $\tau\tau\tau\tau$ and $\tau\tau\mu\mu$ nor the recent
$\gamma\gamma g g$ final states are expected to constrain the displayed parameter plane at the
HL-LHC, based on this extrapolation from the current LHC data.

The $4\mu$ final state covers low masses of a few GeV. However, $\sin\theta$ is in part of this mass
range already strongly constrained by rare $B$-decays. Furthermore, the $4\mu$ bounds at
$13\tev$~\cite{CMS-PAS-HIG-16-035,Aaboud:2018fvk} are reported for a certain model only, namely dark
gauge bosons, a 2-Higgs doublet model with an additional singlet (2HDM+S) or the NMSSM for a fixed
value of $\tan\beta$, making the translation to other models less straightforward.  A
model-independent presentation of the updated bounds in all channels would be helpful.

As a conclusion, the only direct channels evaluated here that have the potential to constrain the
displayed parameter space at the HL-LHC, are $bbll$, $l=\tau$, $\mu$ and $4b$ in $Vh$ production due
to the sizeable branching fractions. However, the strongest vertical asymptote is at $\mphi>26\gev$
whereas Higgs coupling fits at the HL-LHC can exclude $\mphi>20\gev$. Hence at the HL-LHC, there is
no direct search for $h\to\phi\phi\to F$ that is stronger than the indirect HL-LHC bound.

A different picture emerges at lepton colliders running at $\sqrt{s}=240\gev$ for which projections
of the upper bound on exotic Higgs branching ratios in particular in hadronic final states have been
worked out in Ref.~\cite{Liu:2016zki}. Such signatures are very hard to distinguish from background
at hadron colliders, but can be promising at lepton colliders.  Taking CEPC with $5\iab$ as a
benchmark, a strong bound of $\br(h\to\phi\phi\to4b)<3\times10^{-3}$ for
$10\gev\leq\mphi\leq 60\gev$ has been obtained. The resulting exclusion contour in the
$(\mphi, \sin^2\theta)$ plane is shown in Fig.~\ref{fig:DirIndir} (orange, dotted), excluding
$\mphi>11\gev$.  Hence, this direct search is expected to constrain the relaxion parameter space
more strongly than the bound derived from the anticipated precision of the Higgs couplings, see the
corresponding CEPC contour in Fig.~\ref{fig:indirectprobes} that excludes $\mphi>16\gev$.  Likewise,
for the FCCee and the ILC at the same energy, a similar bound on $h\to4b$ is expected according to
their luminosities.  In contrast, it is not easily transferable to CLIC due to its higher energy far
above the $Zh$ threshold. A dedicated analysis would be necessary and useful to determine the reach
of CLIC via these hadronic search channels of exotic Higgs decays.
\begin{table}
 \begin{center}
 \resizebox{\columnwidth}{!}{%
  \begin{tabular}{|c||c|c||c|c|c|c||c|}
  \hline
  $F$	&	exp.	&	Ref.	&	$\sqrt{s}$\,[TeV]	&	$\mathcal{L}_{\rm int}\,[\textrm{fb}^{-1}]$ 	&	$\mphi$\,[GeV]	&	comment	&$\mphi^{\rm HL}$\,[GeV]	\\ \hline \hline
$bb\tau\tau$	&	CMS	&	\cite{CMS:2018lqr}	&	13	&	35.9	&	15-60	&		&	26	\\ \hline
\multirow{2}{*}{$bb\mu\mu$}	&	CMS	&	\cite{Khachatryan:2017mnf}	&	8	&	19.7	&	15-62.5	&		&	27	\\
	&	ATLAS	&	\cite{Aaboud:2018esj}	&	13	&	36.1	&	20-60	&		&	30	\\ \hline
$\tau\tau\mu\mu$	&	CMS	&	\cite{CMS:2018wii}	&	13	&	35.9	&	15-62.6	&		&	-	\\ \hline
$4\tau$	&	CMS	&	\cite{Khachatryan:2017mnf}	&	8	&	19.7	&	5-15	&		&	-	\\ \hline
\multirow{2}{*}{$4\mu$}	&	CMS	&	\cite{CMS-PAS-HIG-16-035}	&	13	&	2.8	&	0.25-8.5	&	NMSSM, $\gamma_D$	&	\multirow{2}{*}{-}	\\
	&	ATLAS	&	\cite{Aaboud:2018fvk}	&	13	&	2.8	&	1-2.5, 4.5-8	&	2HDMS, $Z_D$	&		\\ \hline
\multirow{2}{*}{$4b$}	&	\multirow{2}{*}{ATLAS}	&	\multirow{2}{*}{\cite{Aaboud:2018iil}}	&	\multirow{2}{*}{13}	&	\multirow{2}{*}{36.1}	&	\multirow{2}{*}{20-60}	&	$Zh$	&	27	\\
	&		&		&		&		&		&	$Wh$	&	29	\\ \hline
$\gamma\gamma gg$	&	ATLAS	&	\cite{Aaboud:2018gmx}	&	13	&	36.7	&	20-60	&	VBF	&	-	\\ \hline
  \end{tabular}
  }
  \caption{Summary of the implications of exotic Higgs decay searches on the relaxion parameter
    space. The columns show the final state $F$ of the search channel $pp\to h\to\phi\phi\to F$; the
    experiment (ATLAS, CMS) with reference;
    the data set collected at a center-of-mass energy $\sqrt{s}$ with an
    integrated luminosity of $\mathcal{L}_{\rm int}$, the mass range of $\mphi$ probed by the specific channel,
    comments (on the production mode and model-dependence of some bounds);
    estimate of the asymptotically vertical upper bound on $\mphi$ of the HL-LHC projection (if any).
    For comparison, CEPC is expected to exclude $\mphi\geq 11\gev$ via
    $h\to4b$ (for the corresponding BR, see Ref.~\cite{Liu:2016zki}). }
 \label{tab:exoH}
 \end{center}
\end{table}

\subsubsection{Production at the LHC}
\label{sec:direct-probesLHC}

Similarly to the Higgs, the dominant production modes for the relaxion at the LHC are
\begin{itemize}
\item gluon fusion: $ p  p \to \phi $ 
\item relaxion strahlung: $ p  p \to Z \phi,~W\phi$ 
\item $\left\lbrace t\bar t,~b\bar b \right\rbrace$-associated production:
  $ p p \to\left\lbrace t\bar t \phi,~b\bar b\phi \right\rbrace $
\item vector boson fusion (VBF): $p p \to \phi j j$
\end{itemize}
In the left panel of Fig.~\ref{fig:prod_decay} we present (as solid lines) the production cross
sections $\sigma(pp\to X)$ depending on $\mphi$ in different channels $X=\phi$ (gluon fusion),
$W\phi, Z\phi, t\bar t\phi$, $b\bar b\phi$ and $\phi jj$ (VBF) for $\sin^2\theta=1$. In all
processes (beside the resonant $\phi$ production), we require a minimal transverse momentum of the
relaxion of $p_T(\phi)>20\gev$ and evaluate the cross sections at $\sqrt{s}=13\tev$.  The gluon
fusion cross section was calculated at ${\rm N^3LO}$ and with resummation up to ${\rm N^3LL}$ using
\texttt{ggHiggs v3.5}~\cite{Ball:2013bra,Bonvini:2014jma,Bonvini:2016frm,Ahmed:2016otz}.  All other
cross sections for processes at the HL-LHC were obtained with
\texttt{MadGraph5\_aMC}~\cite{Alwall:2014hca} at NLO.  Compared to the production cross section of
the Higgs with a mass of $m_h=125\gev$, which is of the order of pb, there is at most an enhancement
of two orders of magnitude (reached only for $W\phi$ at the lowest considered $\mphi$) for
$\sin^2\theta=1$ and setting all \cp{}-odd couplings $\tilde c_i=0$. The LEP search, however,
already constrains $\sin^2\theta \lesssim 10^{-2}$ in the considered mass range. As a consequence,
the $\phi$ searches need to be targeted at cross sections smaller than the SM values of the
$125\gev$ Higgs.

Regarding the decay modes, the right panel of Fig.~\ref{fig:prod_decay} shows the branching ratios
of $\phi$ for only \cp{}-even couplings. The leptonic channels, \textit{i.e.}~$ \phi \to \tau \tau$,
$ \mu \mu$ have a similar branching ratio for $\mphi>2m_b$ as for the 125\,GeV Higgs and therefore
the expected rates are small. The other clean channel, $\phi\to\gamma \gamma$, is suppressed by a
factor of 100 to 2 for $\mphi$ between 5 and 100\,GeV, respectively, compared to the BR at a mass of
125\,GeV due to the $\mphi^3$-dependence.  Hence, an observation of this final state might only
become feasible if this partial width is enhanced by the \cp{}-odd coupling
$\tilde c_{\gamma\gamma}$.  While below the $b\bar b$ threshold the decay into $c\bar c$ is
dominant, ${\rm BR}(\phi\to b\bar b)$ approaches 1 where kinematically allowed. However, hadronic
final states of low-mass resonances pose a severe challenge. CMS has already performed a vector
resonance search with hadronic final states, yet only for $\mphi>50\gev$. Further experimental
efforts focused on lower masses would be helpful. Regarding the status at the LHC, CMS searched for
pseudoscalars in the $25\gev$--$80\gev$ mass range produced in association with $ b \bar b$ in the
19.7\;fb$^{-1}$ data set from the 8\;TeV run (see Ref.~\cite{Khachatryan:2015baw} for
$ \tau^+ \tau^{-}$ and Ref.~\cite{Sirunyan:2017uvf} for $\mu^+ \mu^{-}$decay modes). Assuming a
similar efficiency for scalars the present bound is weaker than the one from LEP.  Moreover, a naive
rescaling by the increased luminosity at HL-LHC does not improve on the LEP bound
either. Nonetheless, an improvement beyond the increased luminosity might provide valuable input.

Ref.~\cite{Casolino:2015cza} investigates in a phenomenological study the discovery prospects for a
new pseudoscalar with a mass between 20 and 100\,GeV produced via $t \bar t \phi$ with $\phi$
decaying into $ b \bar b$.  Ref.~\cite{Chang:2017ynj} instead studied the LHC phenomenology of a
light scalar in the same mass range mixed with the Higgs and concluded that the LHC has the
potential to slightly improve the LEP constraints for this channel for $\mphi>80\gev$.  However, the
cross sections to be probed by this analysis are too large to be realized in a relaxion framework.
Recently, the sensitivity of the LHC to light axion-like particles produced in gluon fusion has been
studied for decays into $\gamma\gamma$~\cite{Mariotti:2017vtv} and
$\tau^+\tau^-$~\cite{Cacciapaglia:2017iws}.  Their projected sensitivity could set competitive
bounds for \cp{}-odd couplings, but for \cp{}-even couplings the cross sections within reach
correspond to a mixing angle already probed by LEP.  Potentially more promising production modes are
the associate production with a gauge boson, i.e. $Z\phi$ or $W\phi$, which yields larger cross
sections, as shown in Fig.~\ref{fig:prod_decay}.

So far we have considered the case where the dominant couplings are the ones given by the
Higgs-relaxion mixing.  As we previously discussed, the relaxion also has $\cp$-odd couplings with
the $SU(2)$ gauge bosons which cannot only change the decay pattern, but also the production. In
particular the production $ p p \to W(Z) \phi$ can receive an additional contribution such as
\begin{align} 
  \sigma( p p \to \phi Z )_5 &= 2.7\times 10^4\,\sint^2\,  \fb + 2.2 \left(\frac{1\tev}{f}\right)^2 \left(1.9\,\tilde c_{ZZ}^2
                                                        + 1.1\,\tilde c_{ZZ} \tilde c_{\gamma Z} + \tilde c_{\gamma Z}^2\right)\fb\\
  \sigma( p p \to \phi Z )_{35} &= 1.0\times 10^4 \,\sint^2\,\fb + 2.0 \left(\frac{1\tev}{f}\right)^2 \left(1.8\,\tilde c_{ZZ}^2
                                                         + 1.0\,\tilde c_{ZZ} \tilde c_{\gamma Z} +
                                                         \tilde c_{\gamma Z}^2\right)\fb\\
  \sigma(p p \to \phi W)_5 &= 6.6\times 10^4\,\sint^2\,\fb + 2.7
                                                      \left(\frac{1\tev}{f}\right)^2 \,\tilde c_{WW}^2\,\fb\\
  \sigma(p p \to \phi W)_{35} &= 2.3\times 10^4\,\sint^2\,\fb + 2.4
                                                       \left(\frac{1\tev}{f}\right)^2 \,\tilde c_{WW}^2\,\fb \,,
\end{align}
where the subscripts 5 and 35 denote \mphi{} in GeV.
The partonic cross sections were calculated analytically with \texttt{FeynCalc} version
9.2.0~\cite{Mertig:1990an,Shtabovenko:2016sxi} and folded with the LO parton distribution from
\texttt{NNPDF} version 3.1~\cite{Ball:2017nwa}, interfaced with \texttt{ManeParse} version
2~\cite{Clark:2016jgm}. As a cross check of our calculation, we compared the \cp{}-even part with
the cross section calculated with \texttt{MadGraph5\_aMC}~\cite{Alwall:2014hca} and found agreement.

\begin{figure}[tb]
 \begin{center}
  \includegraphics[width=0.48\textwidth]{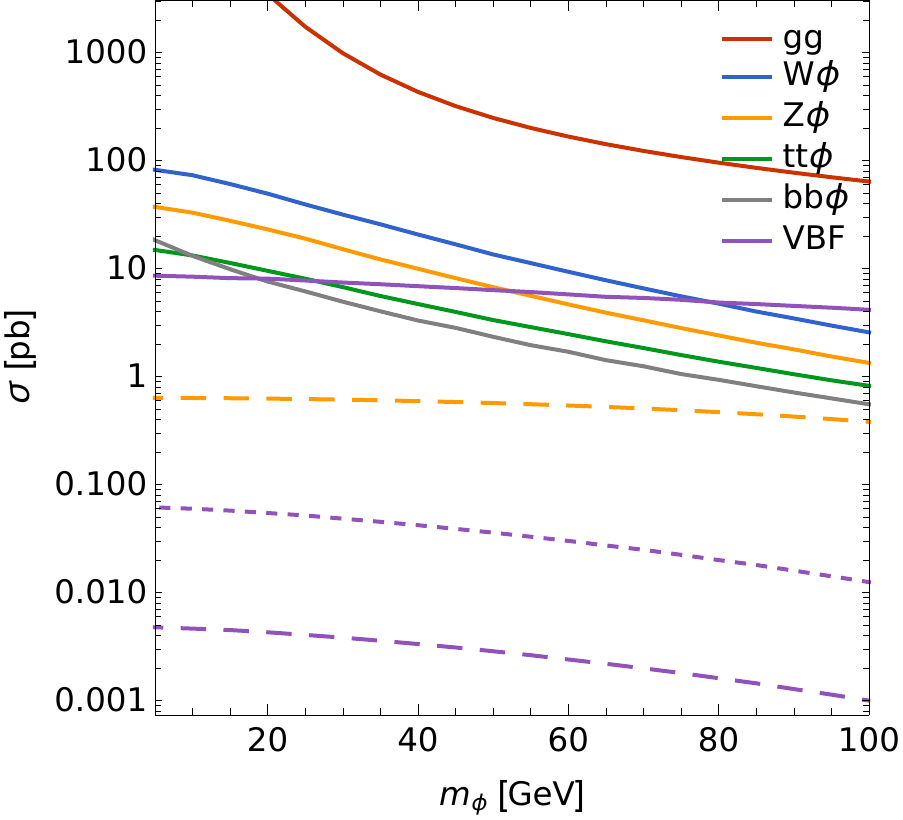}\hfill
  \includegraphics[width=0.48\textwidth]{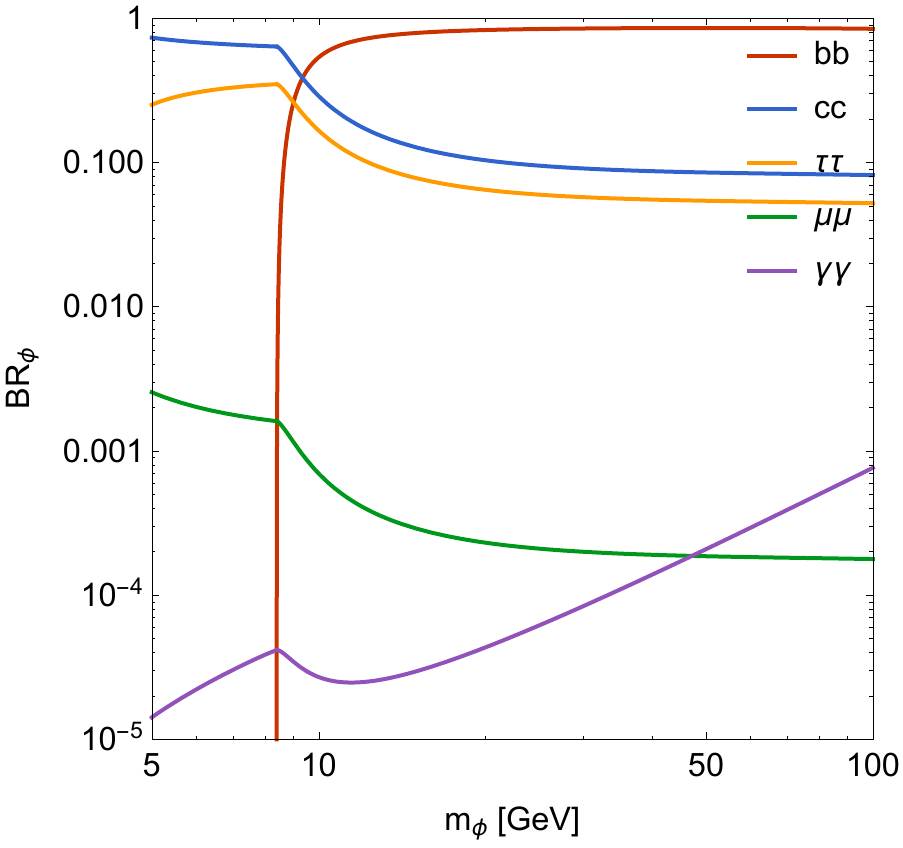}
  \caption{Production and decay of $\phi$ via \cp{}-even couplings for $\sin^2\theta=1$.
    \textbf{Left:} Hadronic cross sections in solid lines, $\sigma(pp\to X)$ at $\sqrt{s}=13\tev$
    for $X=\phi$ (via gluon fusion), $W\phi$, $Z\phi$, $t\bar t\phi$, $b\bar b\phi$ and $\phi jj$
    (via VBF).  Leptonic cross sections in dashed lines, $\sigma(e^+e^-\to Y)$ at $\sqrt{s}=240\gev$
    for $Y=\phi Z, \phi\nu_e\bar\nu_e$ (via $W$-fusion, dotted) and $\phi e^+e^-$ (via $Z$-fusion,
    dashed).  The $\sigma(pp\to\phi)$ via gluon fusion is calculated using \texttt{ggHiggs
      v3.5}~\cite{Ball:2013bra,Bonvini:2014jma,Bonvini:2016frm,Ahmed:2016otz} at N$^3$LO including
    N$^3$LL resummation without a $p_T$-cut.  The remaining hadronic cross sections are obtained
    from \texttt{MadGraph5\_aMC}~\cite{Alwall:2014hca} at NLO with $p_T(\phi)>20\gev$; the leptonic
    VBF cross sections at LO with $p_T(\phi,e^+,e^-)>10\gev$. The leptonic $Z\phi$ cross section was
    analytically calculated with Eq.~(\ref{eq:eeZphiCPeven}), also with $p_T(\phi)>10\gev$.
    \textbf{Right:} Branching ratios
    ${\rm BR}(\phi\to b\bar b,\, c\bar c,\, \tau^+\tau^-, \mu^+\mu^-, \gamma\gamma)$.  }
   \label{fig:prod_decay}
 \end{center}
\end{figure}

\subsubsection{Production at electron colliders}
\label{sec:direct-probeseC}

Due to the clean environment, lepton colliders are not only able to explore the relaxion parameter
space via precision measurements as discussed in the previous section, but also via direct relaxion
production.

When considering the \cp{}-even coupling via mixing with the Higgs, there are two main production
channels for relaxions at lepton colliders, $\phi Z$ and $\phi\nu_e \bar\nu_e$ via $W$-fusion. The
cross section for the third production mode, $\phi e^+ e^-$ via $Z$-boson fusion, is about one order
of magnitude smaller than that of $W$-fusion and will therefore only play a negligible role. For
$\phi Z$ associated production, the polarized cross section is given by
\begin{align}
  \label{eq:eeZphiCPeven}
  \sigma_{\phi Z} &= \frac{\pi\alpha^2 \,\sint^2}{24 s}
                    \frac{\mathcal{P}^2_{P^+P^-}}{\cwsq \swsq}
                    \frac{\left[\lambda_{\rm K}(\sqrt{s},m_Z,m_\phi) + 12m_Z^2 s\right]
                    \sqrt{\lambda_{\rm K}(\sqrt{s}, m_Z,m_\phi)}}
                    {s\left[\left(s-m_Z^2\right)^2+m_Z^2\Gamma_Z^2\right]}\,,
\end{align}
where $P^\pm$ denote the positron and electron polarization, respectively, and
\begin{align}
  \label{eq:polfunc}
  \begin{split}
  \mathcal{P}_{RL}&=-\frac{2\sw}{\cw}\\
  \mathcal{P}_{LR}&=\frac{1-2\swsq}{\sw\cw}\\
  \mathcal{P}_{RR}&=\mathcal{P}_{LL}=0\,,    
  \end{split}
  \end{align}
and $\lambda_{\rm K}$ is given in Eq.~\eqref{eq:mass-difference}.
In the limit $s\gg m_Z \gg \mphi$, the expression simplifies to
\begin{align}
  \label{eq:eeZphiCPevenLimit}
  \sigma_{\phi Z} &\to \frac{\pi\alpha^2 \,\sint^2}{24 s}
    \frac{\mathcal{P}^2_{P^+P^-}}{\cwsq
    \swsq}\left[1+\frac{11 m_Z^2}{s} + \mathcal{O}\left(\frac{\mphi^2}{s}, \frac{m_Z^4}{s^2}\right)\right]\,.
\end{align}
For $\sqrt{s}=240\gev$ and $\mphi=0$,
the following approximate numerical values are obtained from the full expression in Eq.~\eqref{eq:eeZphiCPeven}
\begin{align}
   \label{eq:eeZphiCPevenapprox}
     \sigma_{\phi Z}                   &\approx\left.
                                          \begin{cases}
                                            1.6 & \text{for } P^+P^-=LR\\
                                            1.0 & \text{for } P^+P^-=RL
                                          \end{cases}
                                                  \right\}
                                                 \pb \,\sint^2\,.
\end{align}

The existing limits on this process from LEP2 (ALEPH, DELPHI, L3, OPAL) \cite{Schael:2006cr} were
presented in Ref.~\cite{Flacke:2016szy}.  Our approach here is to estimate the reach of future
lepton colliders running at the $Z$-pole or above by rescaling LEP2 bound by the square root of the
ratio of luminosities.  For the FCCee we assume a luminosity of $10\iab$ at a center-of-mass energy
of $\sqrt{s}=240\gev$.  When extrapolating the LEP2 limits, we neglect the difference in cross
section due to the different center-of-mass energies within the various LEP2 runs (192--202\,GeV)
and between LEP2 and the FCCee.  The $\phi Z$ cross section at $\sqrt{s}=240\gev$ is by a factor of
about 2 smaller than at $\sqrt{s}=\mathcal{O}(200\gev)$.  Yet, at the same time, also the cross
sections of the background processes decrease so that our rescaling by luminosity ratios only is
justified as a rough approximation.

For lighter $\phi$-masses it is more promising to consider the three-body decay $Z\to \ell\ell\phi$
at the run at the $Z$ pole. The TeraZ limit is rescaled from the LEP1 (L3)
limit~\cite{Acciarri:1996um} by a factor of
$\sqrt{\textrm{N}_Z^\textrm{L3}/\textrm{N}_Z^\textrm{TeraZ}}$ where
$\textrm{N}_Z^\textrm{L3}=4.4\times 10^6$ and $\textrm{N}_Z^\textrm{TeraZ} = 10^{12}$ are the number
of $Z$ bosons at the respective collider/experiment.  These estimates are presented by the green
dashed lines in Fig.~\ref{fig:DirIndir}.\par
    
This is a conservative extrapolation of cut-and-count analyses and will be certainly outperformed by
modern analyses. Compared to the LEP analyses they will benefit from improved flavour-tagging
techniques, better detector design and also the advances in computational methods,
\textit{i.e.}~usage of machine-learning tools. The final states might be different from those
considered in the LEP analyses when taking $\cp$-odd couplings to photons or to leptons into
account.  In order to enable an estimate of the potential of these machines independent of the
relaxion decay modes, we show in Fig.~\ref{fig:prod_decay} as dashed lines the unpolarized cross
section $\sigma(e^+e^-\to Z\phi)$ at $\sqrt{s}=240\gev$ for $\sin^2\theta=1$. The lower cross
section with respect to those at the LHC will be compensated by the clean environment, the possible
enhancement by polarization as well as the large planned luminosity, making future lepton colliders
powerful machines for $\phi$ production.

The production of a light scalar $\phi$ with $\mphi>20\gev$ in association with a $Z$ was studied in
the context of the ILC in Ref.~\cite{Drechsel:2018mgd}. In particular, for a luminosity of
$\mathcal{L}=2000\ifb$ and with polarized beams, the LEP2 bounds can be significantly
improved. Their result is shown as the green dotted line in Fig.~\ref{fig:DirIndir}.

The second relevant production mode is $W$-fusion leading to a $\phi\nu_e\bar\nu_e$ final state. The
cross section at $\sqrt{s}=240\gev$ with $p_T(\phi)>10\gev$ is calculated with
\texttt{MadGraph5\_aMC} yielding
\begin{align}
  \sigma(e^+ e^- \to \phi \nu_e  \bar \nu_e)_{5} &= 61\,s_{\theta}^2 \fb\\
  \sigma(e^+ e^- \to \phi \nu_e  \bar \nu_e)_{35} &= 45\,s_{\theta}^2 \fb\,,
\end{align}
where the subscripts 5 and 35 denote \mphi{} in GeV. This process played a marginal role at LEP, but
it is important to be included for instance in searches targeting missing energy final states. For
comparison, the cross sections for $Z$-fusion leading to the $\phi e^+ e^-$ final state are
\begin{align}
  \sigma(e^+ e^- \to \phi e^-  e^+)_{5}  &= 4.8\,s_{\theta}^2 \fb\\
  \sigma(e^+ e^- \to \phi e^-  e^+)_{35} &= 3.6\,s_{\theta}^2 \fb\,.
\end{align}
These values are calculated with an additional cut on the transverse momentum of the two leptons of
10\gev{} and the pseudorapidity $\eta<2.5$.  They are about one order of magnitude smaller than the
corresponding cross sections for $W$-fusion.

\subsubsection{Probing the \cp{}-odd couplings and the relaxion CPV nature}
\label{sec:probing-cp-odd}

We have briefly discussed above the possible (CPV) contributions of the relaxion to the electron EDM
that is proportional both to its \cp{}-even and -odd coupling.  Here we consider the potential to
probe the unique \cp{} properties of the relaxion from processes involving its production at lepton
colliders.  We first discuss the case of measurements involving $\phi Z$ final states, and then move
to examine those with $\phi \gamma\,.$ As for the former channel, the production from mixing,
\textit{i.e.}~through the \cp{}-even vertex, is dominated by the relevant coupling to $Z$ while the
latter is loop-induced and is described by a dimension-five operator in the effective field theory.
This is in contrast with the pseudoscalar interactions that are always induced by dimension-five
operators.

\paragraph{Production via \cp{}-odd couplings}
As for the $\phi Z$ final state we have already discussed above the contribution from mixing and
thus show now the production cross section via \cp{}-odd couplings:
\begin{align}
  \label{eq:eeZphiCPodd}
  \begin{split}
    \tilde\sigma_{\phi Z} &=\frac{\alpha}{48 \left(4\pi f\right)^2
      s}\lambda_{\rm K}(\sqrt{s},m_Z, m_\phi)^{3/2}\times\\ 
          &\quad\left[\frac{\tilde c_{ZZ}^2}{\left(s-m_Z^2\right)^2
        +m_Z^2\Gamma_Z^2}\mathcal{P}_{P^+P^-}^2 
        + \frac{4\, \tilde c_{\gamma Z}^2}{s^2}
        + \frac{4\,\tilde c_{\gamma Z}\,\tilde
        c_{ZZ}\,\left(s-m_Z^2\right)}{s\,\left[\left(s-m_Z^2\right)^2+m_Z^2\Gamma_Z^2\right]}\mathcal{P}_{P^+P^-}\right],
  \end{split}
 \end{align}
 where the polarization factors are the ones given in Eq.~\eqref{eq:polfunc}, and for
 $s\gg \mZsq\gg \mphisq$, it simplifies to
\begin{align}
  \label{eq:eeZphiCPoddLimit}
  \tilde\sigma_{\phi Z}&\to\frac{\alpha\,\left(2\,\tilde c_{\gamma Z} + \tilde c_{ZZ}
    \mathcal{P}_{P^+P^-}\right)^2}{48 \left(4\pi f\right)^2}\left[1- \frac{m_Z^2}{s}\frac{6\, \tilde c_{\gamma Z} + \tilde
    c_{ZZ}\mathcal{P}_{P^+P^-}}{2\,\tilde c_{\gamma Z} + \tilde c_{ZZ}\mathcal{P}_{P^+P^-}} +\mathcal{O}\left(\frac{\mphi^2}{s},\frac{m_Z^4}{s^2}\right)\right]\,.
\end{align}
For $\sqrt{s}=240\gev$ and $\mphi=0$, Eq.~(\ref{eq:eeZphiCPodd}) yields
\begin{align}
  \label{eq:eeZphiCPoddapprox}
   \tilde\sigma_{\phi Z} &\approx 1.0\fb\left(\frac{\textrm{TeV}}{f}\right)^2
                                          \begin{cases}
                                            0.60\, \tilde c_{ZZ}^2 + 1.6\, \tilde c_{ZZ} \tilde c_{\gamma Z} + \tilde c_{\gamma Z}^2 & \text{for } P^+P^-=LR\\
                                            0.39\, \tilde c_{ZZ}^2 - 1.3\, \tilde c_{ZZ} \tilde c_{\gamma Z} + \tilde c_{\gamma Z}^2 & \text{for } P^+P^-=RL\\
                                          \end{cases}\,.
\end{align}
These values are about two orders of magnitude smaller than the $Z\phi$ cross section from the
\cp{}-even interaction given in Eq.~\eqref{eq:eeZphiCPevenapprox}, for $\tilde c_i \sim
\sint$. Therefore the \cp{}-odd interaction only plays a significant role in this process when the
mixing angle is much smaller than the $\tilde c_i$ for $f=1$\,TeV.  Allowing for both \cp{}-even and
-odd couplings to be present, a measurement of the total cross section can only constrain one
(quadratic) combination of these.  However, an angular analysis of the distributions of the $Z$
decay products involving an up-down imbalance is sensitive to the \cp{} asymmetry of the decay (see
\textit{e.g.}~Ref.~\cite{Delaunay:2013npa}). Consequently, one can in principle probe both the
\cp{}-even and -odd components of the cross section.

In addition to the $Z \phi$ production we also consider the $ \gamma \phi$ production. We first note
that for the latter the mixing contributions only arise at the one-loop level and thus this channel
is much less sensitive to these when compared to the $Z \phi$ channel. For simplicity we therefore
focus on the \cp{}-odd couplings with the corresponding cross section given by
\begin{equation}
  \label{eq:eeGammaphiCPodd}
  \tilde\sigma_{\phi\gamma} =\frac{\alpha\,\left(s-m_\phi^2\right)^3}{48\,s\,\left(4\pi
      f\right)^2}\left[
    \frac{\tilde c_{\gamma Z}^2
      \mathcal{P}_{P^+P^-}^2}{\left(s-m_Z^2\right)^2 + m_Z^2\Gamma_Z^2} +
    \frac{4\,\tilde c_{\gamma\gamma}^2}{s^2} + \frac{4\, \tilde c_{\gamma Z}
      \tilde c_{\gamma\gamma} \left(s-m_Z^2\right)\mathcal{P}_{P^+P^-}
    }{s\,\left[\left(s-m_Z^2\right)^2+m_Z^2\Gamma_Z^2\right]}\right]\,,
\end{equation}
where the polarization factors $\mathcal{P}_{P^+P^-}$ are given in Eq.~\eqref{eq:polfunc}. In the
limit $s\gg m_Z^2 \gg m_\phi^2$ it simplifies to
\begin{align}
  \label{eq:eeGammaphiCPoddLimit}
  \tilde \sigma_{\phi\gamma}&\to \frac{\alpha\,\left(2\,\tilde
                              c_{\gamma\gamma} + \tilde c_{\gamma Z}
                              \mathcal{P}_{P^+P^-}\right)^2}{48\left(4\pi f\right)^2}\left[1+\frac{m_Z^2}{s}\frac{2\,\tilde
                              c_{\gamma Z} \mathcal{P}_{P^+P^-}}{2\,\tilde
                              c_{\gamma\gamma} + \tilde c_{\gamma Z}
                              \mathcal{P}_{P^+P^-}} +\mathcal{O}\left(\frac{\mphi^2}{s},\frac{m_Z^4}{s^2}\right)\right]\,,
\end{align}
which agrees with the result presented in Ref.~\cite{Bauer:2017ris} within their assumptions. For
$\sqrt{s}=240\gev$ and $\mphi=0$, Eq.~\eqref{eq:eeGammaphiCPodd} yields
\begin{align}
  \label{eq:eeGammaphiCPoddapprox}
  \tilde \sigma_{\phi\gamma}&\approx 1.6\fb \left(\frac{\textrm{TeV}}{f}\right)^2
                              \begin{cases}
                                0.6\, \tilde c_{\gamma Z}^2 + 1.6\, \tilde c_{\gamma\gamma} \tilde c_{\gamma Z} + \tilde c_{\gamma \gamma}^2 & \text{for } P^+P^-=LR\\
                                0.39\, \tilde c_{\gamma Z}^2 - 1.3\, \tilde c_{\gamma\gamma}
                                \tilde c_{\gamma Z} + \tilde c_{\gamma \gamma}^2 & \text{for }
                                P^+P^-=RL
                              \end{cases}\,.
\end{align}

Finally, rare $Z$ decays into $\phi\gamma$ involve $\tilde c_{\gamma Z}$. For this partial width we
obtain
\begin{align}
 \widetilde\Gamma(Z\to\phi\gamma ) = \tilde c^2_{\gamma Z}\frac{m_Z^3}{96 \pi\left(4\,\pi f\right)^2}\left(1-\frac{m_\phi^2}{m_Z^2}\right)^3\,
 \label{eq:Zgammaphi}
\end{align}
in agreement with \textit{e.g.}~Ref.~\cite{Bauer:2017ris}.

\paragraph{Constraining the \cp{}-odd couplings}
\begin{figure}[tb]
  \centering
  \includegraphics[height=169pt]{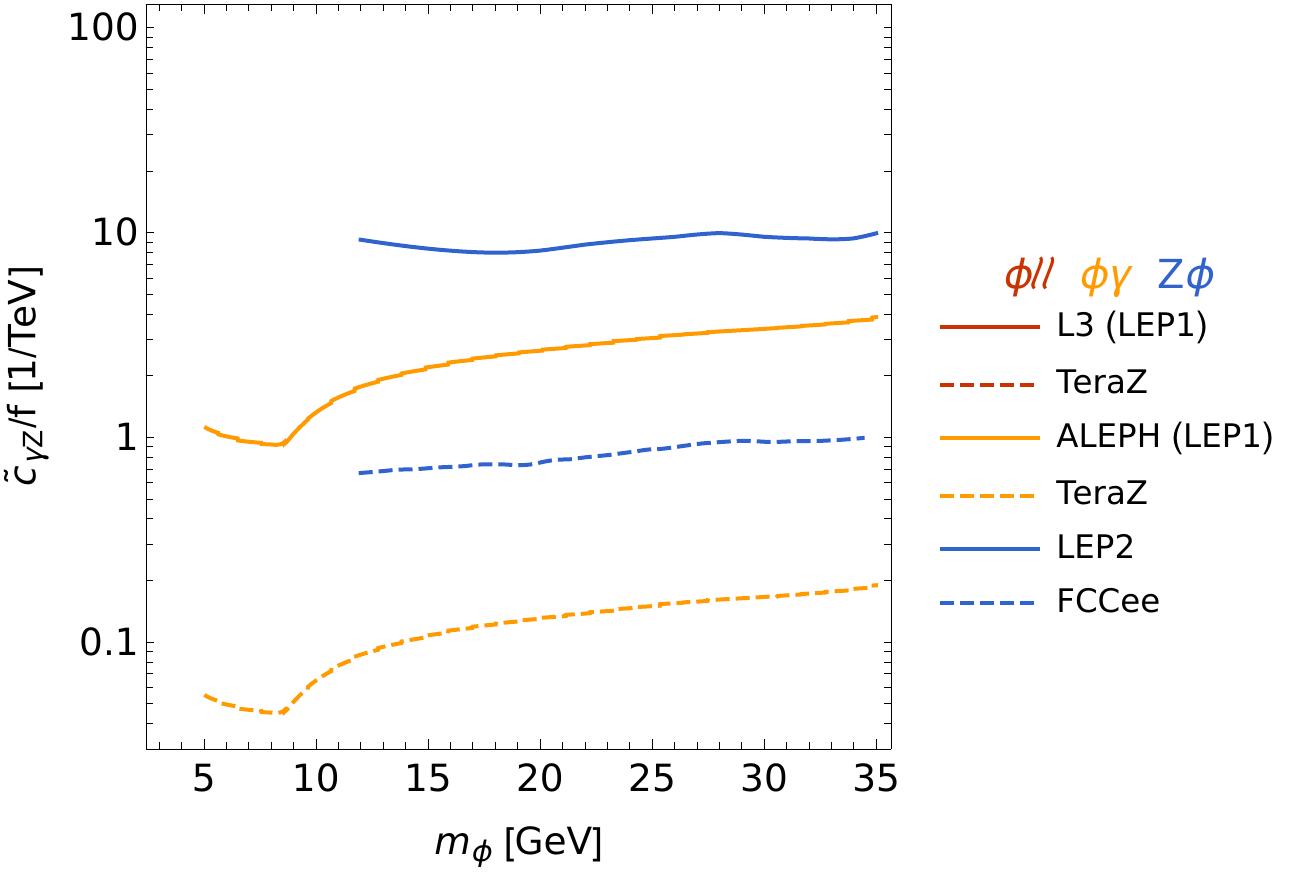}
  \includegraphics[height=169pt]{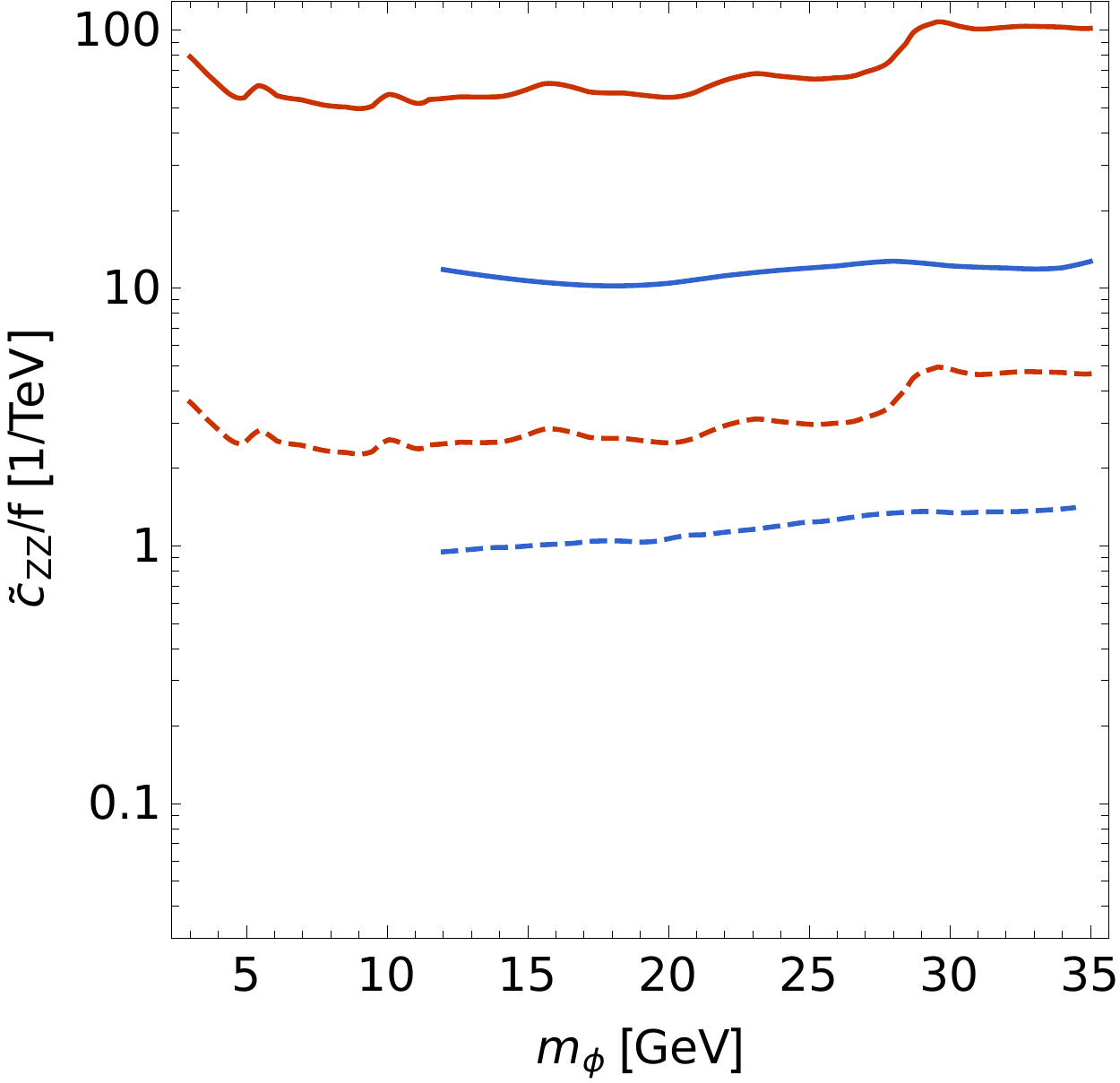}
  \caption{Existing (solid) and projected (dashed) limits on the \cp{}-odd coupling of the relaxion
    to $\gamma Z$ (left) and $Z Z$ (right) as a function of the relaxion mass: $e^+e^-\to Z\phi$
    (blue) at LEP2~\cite{Schael:2006cr} and FCCee. Additionally for $\tilde c_{\gamma Z}/f$:
    $Z\to\gamma \phi$ (orange) at ALEPH~\cite{Barate:346643,ALEPH:2005ab} during LEP1 and at
    TeraZ. Moreover for $\tilde c_{ZZ}/f$: $\Gamma(Z\to \ell\ell \phi)$ (red) at L3 during
    LEP1~\cite{Acciarri:1996um} and TeraZ.}
  \label{fig:cpoddlimits}
\end{figure}

The above cross section prediction for $\tilde \sigma_{\phi Z}$ as well as the decay widths of
$Z\to \ell\ell \phi$ and $Z \to \phi \gamma$ can be used to constrain the \cp{}-odd couplings
$\tilde c_{ZZ}$ and $\tilde c_{\gamma Z}$. We evaluate the implications of the LEP measurements at
and above the $Z$-pole and estimate the sensitivity at TeraZ and the FCCee.

Regarding $Z\to \ell\ell\phi$, we translate the L3 bound from LEP1~\cite{Acciarri:1996um} on
$\sin^2\theta$ shown in Fig.~\ref{fig:DirIndir} into bounds on the \cp{}-odd $\tilde c_{ZZ}$
coupling. For this we require
\begin{align}
 \br_{Z\to\ell\ell\phi}(\sin^2\theta_{\rm max})
 \stackrel{!}{=} \widetilde \br_{Z\to\ell\ell\phi}\left(\left(\tilde c_{ZZ}/f\right)^2_{\rm max}\right)\,,
\end{align}
\textit{i.e.}~ignoring differences in the distributions. Here $\br$ refers to the decay through
mixing where all \cp{}-odd couplings are neglected whereas $\widetilde\br$ refers to the decay based
on the \cp{}-odd coupling $\tilde c_{ZZ}$. For TeraZ, we take the rescaled LEP1 limit on
$\sin^2\theta$.

Likewise we proceed for $Z\to\phi\gamma$ where only $\tilde c_{\gamma Z}$ plays a role. The
strongest bound on $\br(Z\to\phi\gamma)\times\br(\phi\to ff)$ is reported by
ALEPH~\cite{Barate:346643} for $f=\tau$. Here we assume the decay of $Z$ into $\phi\gamma$ via
$\tilde c_{\gamma Z}$ and the decay of the relaxion into $\tau\tau$ via mixing with the BR as shown
in Fig.~\ref{fig:prod_decay}.  Since $\br(\phi\to ff)$ is independent of $\sin^2\theta$, this
approach is valid for sufficiently small $\sin^2\theta$. We rescale the ALEPH limit by the ratio of
$Z$ bosons used in the analysis~\cite{ALEPH:2005ab} to the $10^{12}$ expected at TeraZ.

Concerning $e^+ e^-\to \phi Z$, we analogously require the number of $Z$ bosons, $N_Z$, produced in
this channel via $\tilde c_i$ to be equal to $N_Z$ produced via mixing at the $\sin^2\theta$ that is
maximally allowed by LEP2 or its rescaling to FCCee, respectively. In order to obtain $N_Z$ at LEP2,
we sum over the products of luminosity times cross section evaluated at the energies of the
respective runs~\cite{Assmann:2002th}.

Fig.~\ref{fig:cpoddlimits} shows the current limits and future projections on $\tilde c_{\gamma Z}$
and $\tilde c_{Z Z}$.  While the runs around $200\gev$ set stronger bounds on $\tilde c_{Z Z}$ than
those at $m_Z$, $\tilde c_{\gamma Z}$ is best constrained at the $Z$-pole.

Finally at very high energy lepton colliders like the high-energy stage of CLIC, $W$-boson fusion
becomes an important process to produce axion-like particles~\cite{Buttazzo:2018qqp}. This could be
used to set bounds on $\tilde c_{WW}$.

\subsection{Comparison of direct and indirect probes}
 \begin{figure}[tb]
  \centering
  \includegraphics[width=0.6\textwidth]{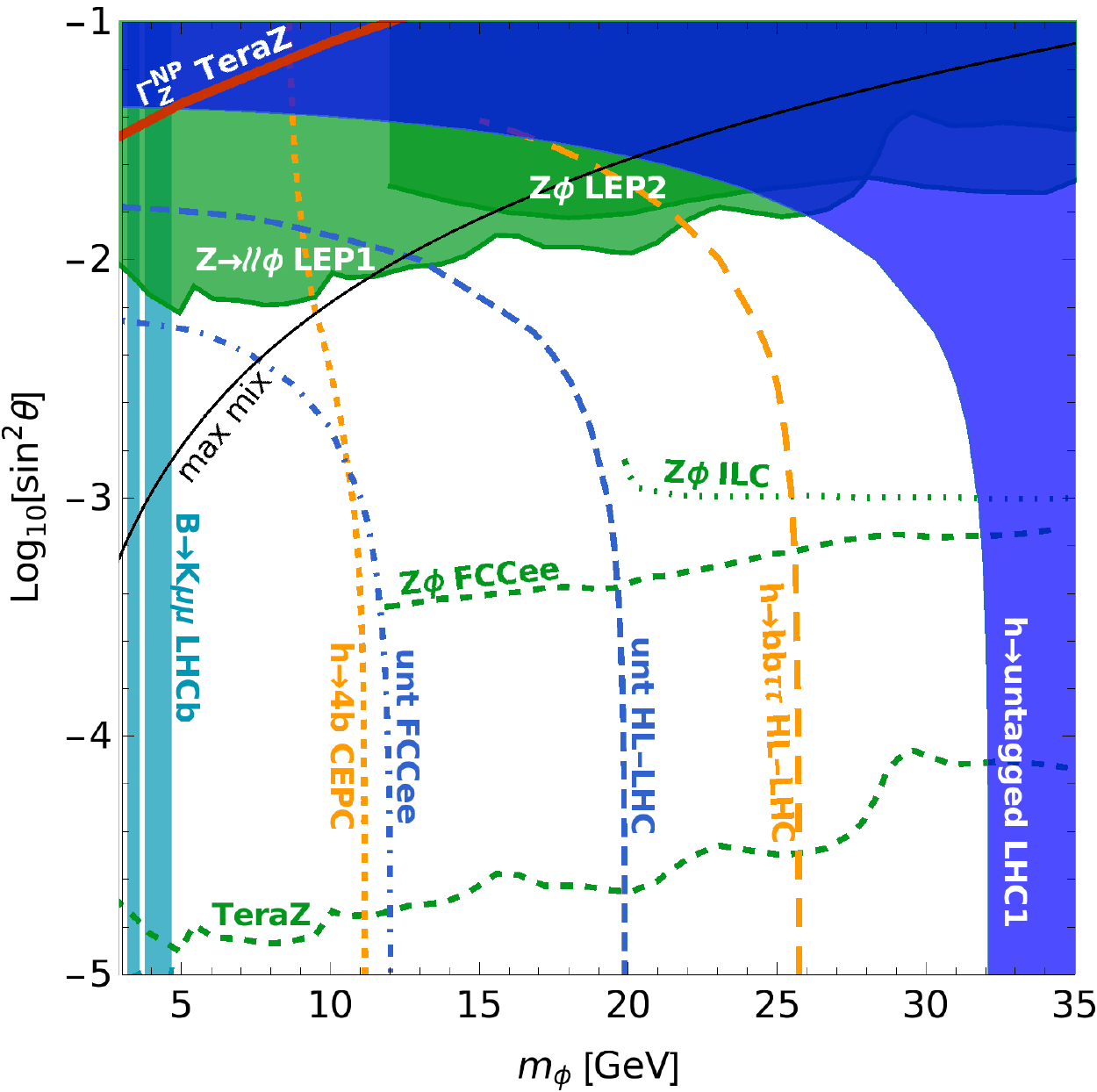}
  \caption{ \emph{Direct} and \emph{indirect} bounds and projections for processes at hadron and
    lepton colliders.  $Z\to Z^*\phi\to \ell\bar \ell \phi$ at LEP1 with
    $\sqrt{s}=M_Z$~\cite{Acciarri:1996um} and $e^+e^-\to Z\phi$ at LEP2 with
    $\sqrt{s}=192$--$202\gev$~\cite{Schael:2006cr}; projections for the same processes at the FCCee
    (green, dashed) running at $\sqrt{s}=M_Z$ with $10^{12}$ $Z$s produced (TeraZ) and
    $\sqrt{s}=240\gev$ with $10\iab$.  Projection for $e^+e^-\to Z\phi$ at the ILC with
    $\lint=2\iab$~\cite{Drechsel:2018mgd} (green, dotted).  Bound from $B^+\to K^+ \mu^+ \mu^-$ at
    LHCb~\cite{Aaij:2012vr,Aaij:2015tna}.  Direct searches for exotic Higgs decays at the HL-LHC in
    the $bb\tau\tau$ channel inferred from Ref.~\cite{CMS:2018lqr} (orange, dashed) and at CEPC with
    $5\iab$ in the $4b$ channel from the BR bound of Ref.~\cite{Liu:2016zki}.  Untagged Higgs decays
    (blue) at the LHC Run-1 (blue area) and projections for the HL-LHC with $3\iab$ (blue, dashed)
    and the FCCee with $10\iab$ (blue, dash-dotted) according to Tab.~\ref{tab:coll_BRuntagged}. The
    NP contribution to the total $Z$-width will be bounded by TeraZ (red). The maximal mixing
    according to Eq.~(\ref{eq:maxmixNatural}) is indicated by the black line.  }
  \label{fig:DirIndir}
\end{figure}

The comparison of the direct and indirect probes in the summary plot of Fig.~\ref{fig:DirIndir}
highlights the complementarity and perspectives in exploring the relaxion framework at colliders.
There are ample regions of parameter space where the relaxion can be discovered both via deviations
in precision measurements of Higgs properties and via direct production.

Focusing on the discovery potential of the future colliders, we notice that the searches for Higgs
decays into a pair of relaxions can probe relaxion masses rather independently of the mixing angle.
Such bounds can be complemented by direct searches in other channels and confirm or rule out a
potential indirect evidence.  Moreover, direct searches are the only way to explore relaxion masses
below 20\,GeV at the (HL-)LHC, providing a strong motivation to try to push the sensitivity below
this mass.  In this context, $\tau$ final states are particularly motivated due to the large
branching ratio in Higgs portal models and a reasonable reconstruction efficiency. At the price of a
lower branching ratio, but with the benefit of a cleaner signature, also other leptonic final states
should be explored, see Fig.~\ref{fig:prod_decay}.

Despite the precision at TeraZ and the expected improvement of the theoretical uncertainty, the NP
contribution to the total $Z$-width will not set new limits beyond the already excluded
range. Nevertheless, direct searches at TeraZ for the decay of $Z\to \phi f\bar f$ do play an
important role in constraining a light $\phi$.

The `collider region' of the relaxion is bounded from below by the LHCb bound of up to 5\,GeV. While
for general scalars any higher masses are relevant, the relaxion window ends at 32\,GeV on account
of the bound set by untagged Higgs decays.

The maximal relaxion-Higgs mixing angle as a function of the mass is limited by
Eq.~(\ref{eq:maxmixNatural}) indicated by the black line. Hence the current bounds hardly probe the
viable region whereas the projections for all of the considered future colliders cut deeply into
this physical parameter region.

\section{Conclusion}
\label{sec:conclusion}

In this paper we evaluated the potential sensitivity of future colliders to light scalars, motivated
by the relaxion framework.  The relaxion phenomenology is driven by two kinds of couplings: The
first kind is in their simplest form identical to those of a \cp{}-even Higgs portal and originates
from the relaxion mixing with the Higgs; the second kind is similar to those of \cp{}-odd
axion-like-particles and typically arises due to an anomalous backreaction sector. We studied
indirect effects and direct production modes both at the HL-LHC and at future lepton colliders,
namely the ILC, CLIC, the CEPC and the FCCee running at the $Z$-pole and above the $Zh$ production
threshold. Our results are applicable to a large class of models beyond the relaxion framework.
Light (pseudo-)scalars can arise in a variety of models, for example Higgs portals, 2HDM with an
additional singlet, and supersymmetric versions thereof, as well as in models with axion-like
particles.

The fact that the relaxion acquires both \cp{}-even and -odd couplings makes it an interesting model
to study, in particular given the powerful capabilities of future colliders.  While focusing mainly
on the implications of the \cp{}-even interactions, we also point out in which channels the
\cp{}-odd couplings can influence the collider phenomenology and derive constraints on them.

Among the \textit{indirect} probes, we evaluated the NP contribution to the total $Z$-width, in this
case from the decay $Z\to f\bar f \phi$, whose current bound from LEP1 will be significantly
improved at TeraZ. Furthermore, we studied the sensitivity to the $h\phi\phi$ coupling via untagged
$h\to\phi\phi$ decays, resulting in strong, though relaxion-specific bounds.

Regarding the \textit{direct} probes, we considered gluon fusion, associated production as well as
explicit searches for $h\to\phi\phi$. We provided analytic expressions for polarized cross sections
at lepton colliders and semi-analytic functions for the processes at the HL-LHC, involving both the
\cp{}-even and -odd couplings.

Our main findings are:
\begin{itemize}
\item We chart the physical parameter region of relaxion models, expressed via the relaxion-Higgs
  mixing angle as a function of the mass. We find that only mixing angles smaller than twice the
  ratio of mass to the Higgs vacuum expectation value can describe actual relaxion models. It is
  quite interesting that the HL-LHC and future colliders are capable to significantly probe this
  "physical relaxion" region, albeit only for a very massive relaxion.
\item The HL-LHC can probe relaxion masses roughly above $20\gev$ by untagged Higgs decays. However,
  its discovery prospects in direct channels via its mixing with the Higgs are limited due to the
  large background for low-mass resonances and suppressed branching ratios for the clean final
  states.
\item Future lepton colliders have the potential to significantly improve on existing limits and
  HL-LHC projections on the parameter space via direct and indirect channels.  In particular masses
  roughly above $10\gev$, and mixing angles above few times $10^{-3}$ will be probed by exotic Higgs
  and exclusive $Z$ decays respectively.
\item The \cp{}-nature of $\phi$ can be deciphered by the interplay of \cp{}-even and -odd
  driven signals. In particular the interplay of collider observables (such as $\phi Z$ and possibly
  $\phi \gamma$ production as well as the angular distribution of decay products) and EDMs is crucial.
\end{itemize}

\section*{Acknowledgments} 
We would like to thank Liron Barak, Roberto Franceschini, Rick S. Gupta, Marumi Kado, Yevgeny Kats, Yosef Nir, Diego Redigolo, Francesco Spano and Tim Stefaniak for useful discussions.
The work of GP is supported by grants from the BSF, ERC,
ISF,  Minerva Foundation,  and  the  Weizmann-UK  Making  Connections Program.
EF is supported by the Minerva Foundation.

\clearpage

\bibliographystyle{JHEP}
\bibliography{RelaxionCollider}
\markboth{}{}
\end{document}
